\documentclass[aps,prc,preprint,groupedaddress,showpacs]{revtex4}

\usepackage{graphicx}
\usepackage{amssymb}

\begin{document}


\title{Hadronic observables from $Au+Au$ collisions at $\sqrt {s_{NN}}=200$ GeV and $Pb+Pb$  
collisions at $\sqrt {s_{NN}}=5.5$ TeV from a simple kinematic model}


\author{T. J. Humanic}
\email[]{humanic@mps.ohio-state.edu}
\affiliation{Department of Physics, The Ohio State University,
Columbus, Ohio, USA}


\date{\today}

\begin{abstract}
A simple kinematic model based on superposition of $p+p$ collisions, relativistic geometry and final-state hadronic rescattering
is used to calculate various hadronic observables in $\sqrt {s_{NN}} = 200$ GeV
$Au+Au$ collisions and $\sqrt {s_{NN}} = 5.5$ TeV $Pb+Pb$ collisions. The model calculations
are compared with experimental results from several $\sqrt {s_{NN}} = 200$ GeV $Au+Au$ 
collision studies. If a short hadronization time is assumed in the model, it is found that this model 
describes the trends of the observables from these experiments
surprisingly well considering the model's simplicity. This also gives more credibility to the
model predictions presented for $\sqrt {s_{NN}} = 5.5$ TeV $Pb+Pb$ collisions. 
\end{abstract}


\maketitle


\section{Introduction}
The experiments at the Relativistic Heavy Ion Collider (RHIC) have produced many interesting studies of hadronic observables from relativistic heavy-ion collisions over the past six or so years. The goal has been to use these observables to characterize the conditions of the early state of matter in heavy-ion collisions so as to be possible signatures of exotic states, such as Quark Matter \cite{rhic1,rhic2,rhic3,rhic4}. Hadronic observables measured at RHIC can be placed into four general categories: spectra, elliptic flow, femtoscopy, and high $p_T$. Examples of observables in each category relevant to the present work are the following: ``spectra'' encompasses rapidity, transverse momentum and transverse mass distributions \cite{phobos1,Adams:2003xp,Adler:2003au}; ``elliptic flow'', characterized by the quantity $V_2$, includes $V_2$ vs. $\eta$ and $V_2$ vs. $p_T$ distributions \cite{Back:2004mh,Adams:2004bi,Ab:2008ed,Adare:2006ti}; ``femtoscopy'' , also known as Hanbury-Brown-Twiss interferometry \cite{hbt1}, includes two-pion correlation studies vs. transverse momentum and azimuthal angle \cite{Adams:2003ra,Adams:2004yc}; and ``high $p_T$'' , which is targeted to be sensitive to jet effects, includes $R_{AA}$ vs. $p_T$ and $dn/d\Delta\phi$ distributions \cite{Adler:2003au,Adler:2002tq}. Models which describe the early stages of the collision after the initial nuclei have passed through each other in terms of partonic degrees of freedom, for example as a cascade or in terms of hydrodynamics, have been successful in describing the experimental systematics of some of these observables in some kinematic ranges, but no single model has thus far succeeded in making an adequate overall description of the systematics of all of these observables  in a wide kinematical range (for a comprehensive overview of comparisons of models with RHIC experiment, see References \cite{rhic1,rhic2,rhic3,rhic4} and references therein).

The goal of the present work is to see how far one can get in describing the experimental systematics of all of the observables mentioned above in a wide kinematical range using a simple kinematic model with hadronic degrees of freedom. In essence the model is, for each heavy-ion collision, a superposition of $p+p$ collisions in the geometry of the colliding nuclei with a proper time for hadronization determining the initial space-time position of each  produced particle, followed by a Monte Carlo hadronic rescattering calculation. The $p+p$ collisions are generated by the PYTHIA code \cite{pythia6.4} at the beam energy of interest. Some of the advantages of using this scheme are clear:
\begin{itemize}
\item One has access to all of the particle types available in PYTHIA.
\item It is conceptually simple: $p+p$ superposition+simple geometry+hadronic rescattering.
\item It has few free parameters -- even the hadronization proper time can be set by a Tevatron study (see below).
\item Jets are automatically included in the model since PYTHIA has jets and thus hadronic observables depending on jets can be studied.
\item The model should be easily scalable via PYTHIA to higher energies such as will be found at the Large Hadron Collider (LHC) thus giving the possibility of making predictions for heavy-ion collisions at those energies.
\end{itemize}

There is no {\it a priori} reason why such an approach should be successful, and in fact there are reasons to think it should be unsuccessful, the most serious one being that it is hard to imagine that hadronic degrees of freedom, rather than partonic degrees of freedom, can be valid soon after the nuclei have passed through each other due to the expected high energy density. This would require a very short hadronization time in these collisions. On this point, it is encouraging that a recent study of pion HBT in Tevatron collisions has shown that a similar model for $p+p$ collisions can explain the $p_T$ and multiplicity dependences for the extracted radius parameters if a very short proper time for hadronization of 0.1 fm/c is assumed \cite{Humanic:2006ib} .

Previous studies using a model similar to this in which hadronic degrees of freedom were assumed in the early stage of the heavy-ion collision followed by hadronic rescattering have been shown to give qualitative agreement with experimental results for some observables  \cite{Humanic:2006a,Humanic:1998a,Humanic:2006b}. Although similar, there are significant differences between those previous studies and the present one:
\begin{itemize}
\item In the old model the initial kinematic state of the hadrons in the collision was parameterized as a thermal distribution in $p_T$ and a Gaussian distribution in rapidity in which the temperature of the thermal distribution and width and mid-rapidity density of the rapidity distribution (as well as the rapidity densities of different particle species) were fixed by comparisons with experiment. In the present model, superposed PYTHIA $p+p$ collisions provide all of the information about the initial kinematic state of the hadrons, including jets which were not present in the old model.
\item In the old model the initial hadronic geometry was taken to be similar to a "Bjorken tube", in that there was no initial expansion in the direction transverse to the beam direction but initial expansion could occur along the longitudinal direction controlled by the hadronization proper time. In the present model, initial expansion is also allowed in the transverse direction to satisfy causality in this picture
(see Eqs. (1) and (2) below).
\item In the previous studies, calculations were done at fixed impact parameter, so comparisons with experimental results which were measured in centrality windows from multiplicity cuts were only qualitative. In the present study the model is run in a "minimum bias" mode in which a distribution of impact parameters is calculated and comparisons with experiments are made using multiplicity cuts to obtain centrality windows equivalent to those from experiment, allowing quantitative comparisons with experiment.
\item In the previous studies hadronic observables were only calculated for the ``soft sector'', i.e. $p_T < 2$ GeV/c due to $p_T$ distributions for hadrons from the old model becoming exponentially larger than experiment for $p_T > 2-3$ GeV/c. It has been found that this behavior was due to an error in the inelastic scattering algorithm in the old model and has been corrected in the present model. Thus, ``hard sector'', i.e. $p_T > 2$ GeV/c, studies are now possible.

\end{itemize}

Still, the only way to determine whether such a radical and simple picture for heavy-ion collisions is valid at all is to compare the results of the model with a range of experimental data. Being such a simple model, the main hope will be to give, at best,  a qualitative description of the trends of the experimental hadronic observables mentioned above. This would already be a useful result since it would help establish the hadronization timescale in RHIC heavy-ion collisions.

To this end, model calculations will be compared with results from the RHIC experiments PHOBOS \cite{phobos1, Back:2004mh}, STAR \cite{Adams:2003xp,Adams:2004bi,Ab:2008ed,Adams:2003ra,Adams:2004yc,Adler:2002tq} and PHENIX \cite{Adler:2003au,Adare:2006ti} for $Au+Au$ collisions at $\sqrt {s_{NN}} = 200$ GeV.
The goal will be to make as quantitative comparisons as possible between model and experiments. Predictions from the model for LHC-energy $Pb+Pb$ collisions at $\sqrt {s_{NN}} = 5.5$ TeV will also be given.

The paper is organized into the following sections: Section II gives a description of the model, Section III presents results of the model for $Au+Au$ collisions at $\sqrt {s_{NN}} = 200$ GeV and comparisons
with RHIC experiments, Section IV presents predictions from the model for
$\sqrt{s_{NN}}=5.5$ TeV $Pb+Pb$ collisions, and Section V gives a summary and conclusions.

\section{Description of the model}
The model calculations are carried out in five main steps: A) generate hadrons in $p+p$  collisions from PYTHIA, B) superpose $p+p$ collisions in the geometry of the colliding nuclei, C) employ a simple space-time geometry picture for the hadronization of the
PYTHIA-generated hadrons,   D) calculate the effects of final-state rescattering among the hadrons,
and E) calculate the hadronic observables. These steps will now be discussed in more detail.

\subsection{Generation of the $p+p$ collisions with PYTHIA}
\begin{figure}
\begin{center}
\includegraphics[width=100mm]{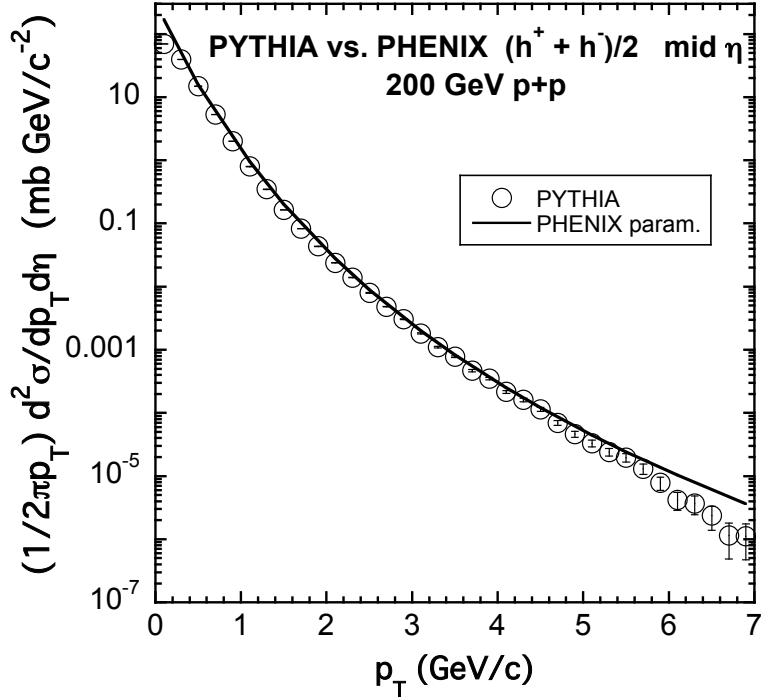} \caption{$d\sigma/dp_T$ from PYTHIA compared
with a PHENIX parameterization for 200 GeV $p+p$ collisions.}
\label{fig1}
\end{center}
\end{figure}

The $p+p$ collisions were modeled with the PYTHIA code \cite{pythia6.4}, version 6.409. The parton distribution functions used were the same as used in Ref. \cite{Humanic:2006ib}. Events were generated
in ``minimum bias'' mode, i.e. setting the low-$p_T$ cutoff for parton-parton collisions to zero (or
in terms of the actual PYTHIA parameter, $ckin(3)=0$) and excluding elastic and diffractive collisions (PYTHIA parameter $msel=1$). Runs were made both with $\sqrt{s}= $ 200 GeV
and 5.5 TeV to simulate RHIC and LHC collisions, respectively. Information saved
from a PYTHIA run for use in the next step of the procedure were the momenta and identities
of the ``direct'' (i.e. redundancies removed) hadrons (all charge states) $\pi$, $K$, $p$, $n$,
$\Lambda$, $\rho$, $\omega$, $\eta$, ${\eta}'$, $\phi$, and $K^*$. These particles were
chosen since they are the most common hadrons produced and thus should have the greatest
effect on the hadronic observables in these calculations.
Although $\Delta$s were included in the rescattering process (see below) during which
they are produced abundantly,
they were
not input directly from PYTHIA since their lifetime is short, i.e. $\approx 1$ fm, and thus
their initial presence was judged to not effect the results significantly. 
Figure \ref{fig1} shows an absolute comparison between the invariant cross section distribution for charged particles from the PYTHIA $p+p$ run (with resonances decayed) used to generate the $Au+Au$ collisions in the present model with a paramerization for 200 GeV $p+p$ collisions from PHENIX \cite{Adler:2003au}. As seen, PYTHIA agrees quite well with the PHENIX parameterization up to about 6 GeV/c, showing that the $p+p$ collisions input into the model are quite reasonable.

\subsection{Superposition of $p+p$ events to simulate heavy-ion collisions}
An assumption of the model is that an adequate job of describing the heavy-ion collision can be obtained by superposing PYTHIA-generated  $p+p$ collisions calculated at the beam $\sqrt s$
within the collision geometry of the colliding nuclei. Specifically, for a collision of impact
parameter $b$, if $f(b)$ is the fraction of the overlap volume of the participating parts of the nuclei such that $f(b=0)=1$ and $f(b=2R)=0$, where $R=1.2A^{1/3}$ and $A$ is the mass 
number of the nuclei, then
the number of $p+p$ collisions to be superposed will be $f(b)A$. The positions of the superposed $p+p$ pairs are randomly distributed in the overlap volume and then projected onto the $x-y$ plane which is transverse to the beam axis defined in the $z$-direction. The coordinates for a particular
$p+p$ pair are defined as $x_{pp}$, $y_{pp}$, and $z_{pp} = 0$. 
The positions of the hadrons produced in 
one of these $p+p$ collisions are defined with respect to the position so obtained of the superposed
$p+p$ collision (see below). 

For fine tuning of the model so as to get slightly better agreement with the PHOBOS rapidity distributions \cite{phobos1}, a lower multiplicity cut on the $p+p$ events which are used in the heavy-ion calculations was made. For $\sqrt {s_{NN}} = 200$ GeV $Au+Au$ calculations, the cut was set to 20, which cut out approximately 26\% of $p+p$ events, and for $\sqrt {s_{NN}} = 5.5$ TeV $Pb+Pb$ calculations, the cut was set to 38 to cut out a similar fraction of $p+p$ events at that energy. It should be emphasized that this was indeed a ``fine tuning'' cut -- if it is not made, the overall results presented below from the model do not change appreciably, showing the robustness of the model calculations to this cut.

\subsection{The space-time geometry picture for hadronization}
The space-time geometry picture for hadronization from a superposed $p+p$
collision located at $(x_{pp},y_{pp})$ consists of the emission of a PYTHIA
particle from a thin uniform disk of radius 1 fm in the $x-y$ plane followed by
its hadronization which occurs in the proper time of the particle, $\tau$. The space-time
coordinates at hadronization in the lab frame $(x_h, y_h, z_h, t_h)$ for a particle with momentum
coordinates $(p_x, p_y, p_z)$, energy $E$, rest mass $m_0$, and transverse disk
coordinates $(x_0, y_0)$, which are chosen randomly on the disk,  can then be written as

\begin{eqnarray}
x_h = x_{pp} + x_0 + \tau \frac{p_x}{m_0} \\
y_h = y_{pp} + y_0 + \tau \frac{p_y}{m_0} \\
z_h = \tau \frac{p_z}{m_0} \\
t_h = \tau \frac{E}{m_0}
\end{eqnarray}

Eqs. (1) and (2) show the initial expansion in the transverse direction now present in the model. 
The simplicity of this geometric picture is now clear: it is just an expression of causality with the
assumption that all particles hadronize with the same proper time, $\tau$. A similar hadronization
picture (with an initial point source) has been applied to $e^+-e^-$ collisions\cite{csorgo}.
For all results presented in this work,  $\tau$ will be set to 0.1 fm/c to be consistent with the results
found in the Tevatron HBT study mentioned earlier \cite{Humanic:2006ib}.

\subsection{Final-state hadronic rescattering}
The hadronic rescattering calculational method used is similar to that
employed in previous studies \cite{Humanic:1998a,Humanic:2006a},
except, as mentioned above, the error found in the algorithm to calculate
inelastic scattering has been corrected. 
Rescattering is simulated with a semi-classical Monte Carlo
calculation which assumes strong binary collisions between hadrons.
Relativistic kinematics is used throughout. The hadrons considered in the
calculation are the most common ones: pions, kaons,
nucleons and lambdas ($\pi$, K,
N, and $\Lambda$), and the $\rho$, $\omega$, $\eta$, ${\eta}'$,
$\phi$, $\Delta$, and $K^*$ resonances.
For simplicity, the
calculation is isospin averaged (e.g. no distinction is made among a
$\pi^{+}$, $\pi^0$, and $\pi^{-}$).

The rescattering calculation finishes
with the freeze out and decay of all particles. Starting from the
initial stage ($t=0$ fm/c), the positions of all particles in each event are
allowed to evolve in time in small time steps ($\Delta t=0.5$ fm/c)
according to their initial momenta. At each time step each particle
is checked to see a) if it has hadronized ($t>t_h$, where $t_h$ is given in
Eq. (4)), b) if it
decays, and c) if it is sufficiently close to another particle to
scatter with it. Isospin-averaged s-wave and p-wave cross sections
for meson scattering are obtained from Prakash et al.\cite{Prakash:1993a}
and other cross sections are estimated from fits to hadron scattering data
in the Review of Particle Physics\cite{pdg}. Both elastic and inelastic collisions are
included. The calculation is carried out to 200 fm/c for
$\sqrt {s_{NN}} = 200$ GeV $Au+Au$ collisions and to 400 fm/c for $\sqrt {s_{NN}} = 5.5$ TeV
$Pb+Pb$ collisions which
allows enough time for the rescattering to finish (as a test, calculations were also carried out for
longer times with no changes in the results). Note that when this cutoff time is reached, all un-decayed resonances are allowed to decay with their natural lifetimes and their projected decay positions and times are recorded.

\begin{figure}
\begin{center}
\includegraphics[width=100mm]{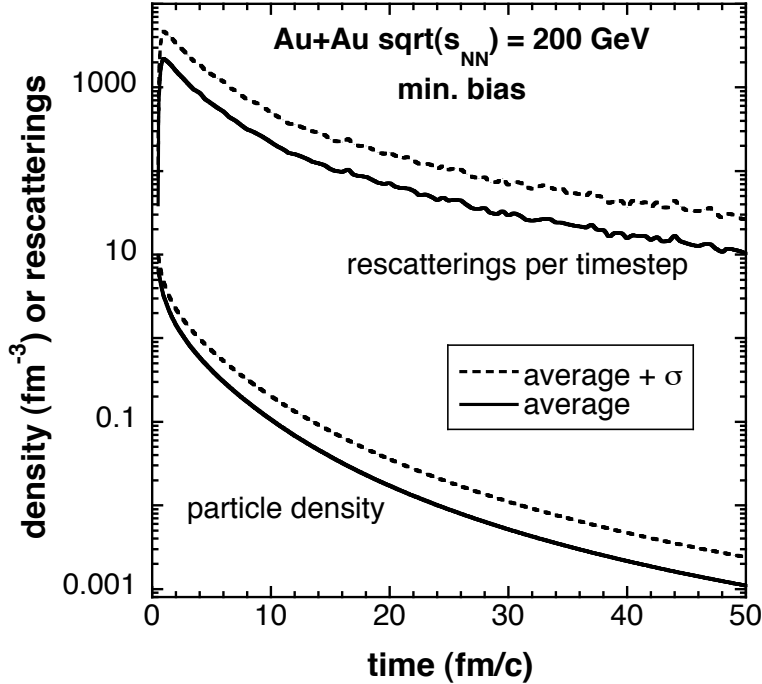} \caption{Time evolution up to 50 fm/c 
of the particle density calculated at mid-rapidity ($-1<y<1$)
and the number of rescatterings per time step from the model for minimum bias 
$\sqrt {s_{NN}}=200$ GeV $Au+Au$ collisions. The solid lines show the average
values of the quantities whereas the dotted lines
show the average $+ \sigma$. }
\label{fig2}
\end{center}
\end{figure}

Figure \ref{fig2} shows the time evolution plotted up to 50 fm/c of the particle density calculated
at mid-rapidity, i.e. in the rapidity range $-1<y<1$,
and the number of rescatterings per time step from the model for minimum bias (see below) 
$\sqrt {s_{NN}}=200$ GeV $Au+Au$ collisions. The solid lines show the average 
values of the quantities whereas the dotted lines
show the average $+ \sigma$ to give an idea of the width of the distribution. 
The density is seen to start out high 
at 6-10 fm$^{-3}$ and then to fall off rapidly with time such that by 4 fm/c the 
density is at or below 1 fm$^{-3}$. The rescattering rate starts small in the first time bin due to
time dilation of the hadronization time and the requirement imposed in the model that particles must hadronize before they can scatter. By the second time step the rescattering rate increases quickly
and then falls off rapidly with time as does the density. The time evolution of the density
will be discussed more later.

The rescattering calculation is described in more detail elsewhere
\cite{Humanic:2006a,Humanic:1998a}. The validity of the numerical
methods used in the rescattering code have recently been studied using
the subdivision method, the results of which have verified that the methods used are 
valid \cite{Humanic:2006b}.

\subsection{Calculation of the hadronic observables}
Model runs are made to be "minimum bias" by having the impact parameters of collisions follow the distribution $d\sigma/db \propto  b$, where $0<b<2R$. Observables are then calculated from the model in the appropriate centrality bin by making multiplicity cuts as done in the experiments, as well as kinematic cuts on rapidity and $p_T$. For the present study, a single 87K event minimum bias run was made from the model for $\sqrt {s_{NN}}=200$ GeV $Au+Au$ collisions which was then used to calculate all of the hadronic observables from the model which are shown  in this work for that system. In this way, a consistent picture of the agreement between the model and experiments can emerge since it is virtually impossible to optimize the model to agree with experiment for a particular observable without spoiling the agreement for others which are calculated from the same run. In the case of the
$\sqrt {s_{NN}}=5.5$ TeV $Pb+Pb$ predictions, a single minimum bias run with 800 events was made.

\section{Results from the model for $Au+Au$ collisions at $\sqrt {s_{NN}}= 200$ GeV and comparisons with RHIC experiments }
Various hadronic observables have been calculated from the 87K minimum bias run from the model mentioned above and are now compared with measurements from RHIC experiments. The observables and experiments with which to be compared are:
\begin{itemize}
\item Spectra -- $dn/d\eta$ (PHOBOS), $dn/dp_T$ (PHENIX), $dn/dm_T$ (STAR)
\item Elliptic flow -- $V_2$ vs. $\eta$ (PHOBOS), $V_2$ vs. $p_T$ charged particles (STAR),
$V_2$ vs. $p_T$ identified particles and $V_2/n_q$ vs. $p_T/n_q$ (PHENIX)
\item HBT -- $\pi\pi$ vs. $\phi$ and $\pi\pi$ vs. $k_T$ (STAR)
\item High $p_T$ -- $R_{AA}$ vs. $p_T$ (PHENIX), $dn/d\Delta\phi$ vs. $\Delta\phi$ (STAR)
\end{itemize}
In the spirit of making as quantitative comparisons as possible between model and experiments, unless explicitly specified otherwise, absolute normalizations are used for the model observables in the plots shown.

\subsection{Spectra}
\begin{figure}
\begin{center}
\includegraphics[width=100mm]{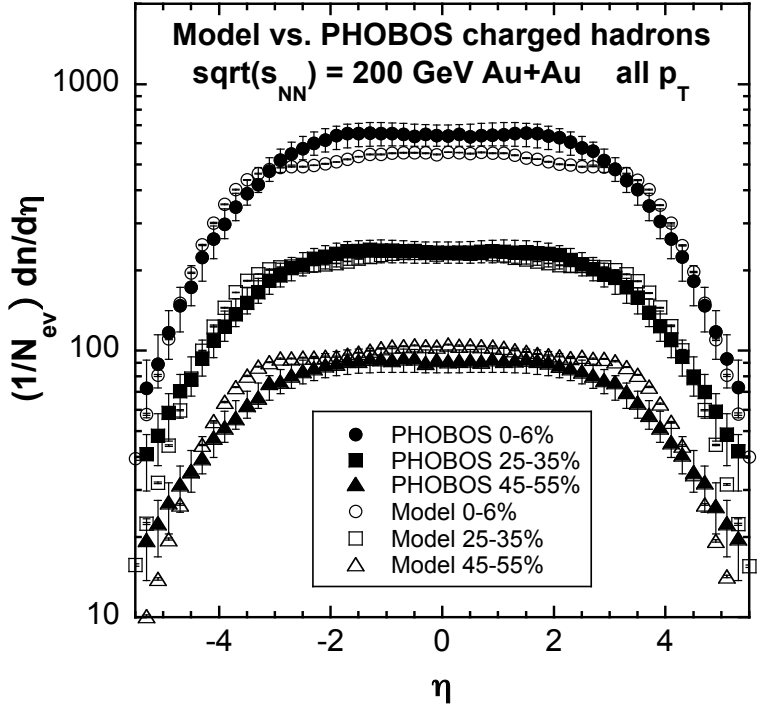} \caption{Rapidity distributions for Model compared
with PHOBOS for several centralilties.}
\label{fig3}
\end{center}
\end{figure}

\begin{figure}
\begin{center}
\includegraphics[width=100mm]{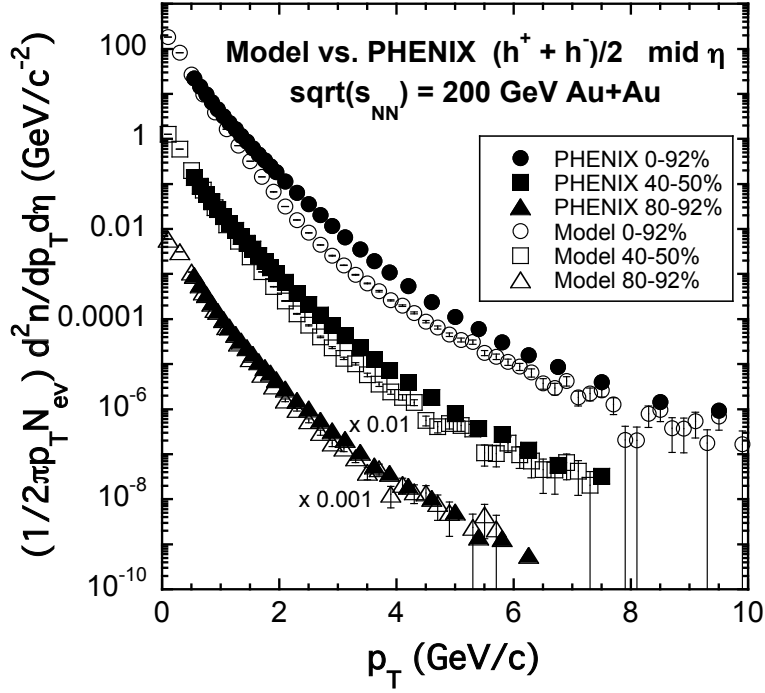} \caption{$p_T$ distribution for charged hadrons
for Model compared with PHENIX for several centralities. The meaning of ``mid-$\eta$''
in this case is that the PHENIX spectra are measured in the range $-0.18<\eta<0.18$, whereas the
model spectra are calculated in the range $-1<\eta<1$ for better statistics. }
\label{fig4}
\end{center}
\end{figure}

\begin{figure}
\begin{center}
\includegraphics[width=100mm]{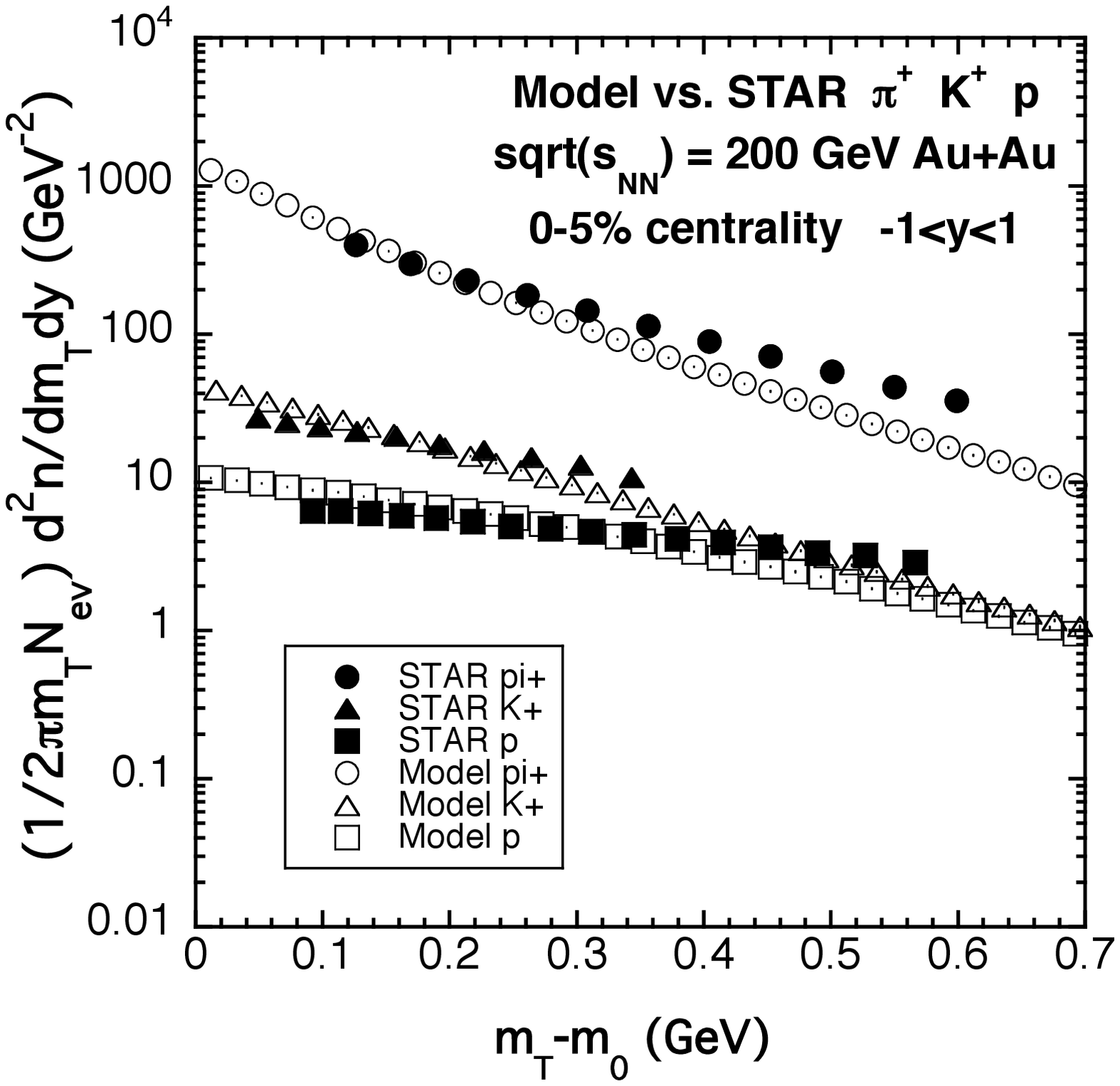} \caption{$m_T$ distributions for
identified particles for Model compared with STAR.}
\label{fig5}
\end{center}
\end{figure}

Figures \ref{fig3}, \ref{fig4}, and \ref{fig5} show model comparisons with PHOBOS \cite{phobos1}, PHENIX \cite{Adams:2003xp}, and STAR \cite{Adler:2003au} for $\eta$, $p_T$, and $m_T$ distributions, respectively. In Figures \ref{fig3} and \ref{fig4} the centrality dependence of charged hadrons is also shown, and in Figure \ref{fig5} the particle species dependence is shown. In Figure \ref{fig3}, the centrality dependence of the rapidity distribution is followed fairly well by the model, although it is seen that the model shapes are slightly broader than experiment and the overall agreement near mid-rapidity is at the 10-15\% level. Since the model is ``isospin averaged'', the model distributions are multiplied by $2/3$ to approximate all charged particles. In Figure \ref{fig4}, the experimental centrality dependence is once again seen to be described reasonably well by the model, especially the absolute scale at low $p_T$. At higher $p_T$, particularly in the minimum bias case, the model under predicts the absolute scale by as much as a factor of four at some points over the $p_T$ range, yet still follows the trend of the data up to the highest $p_T$ shown. To approximate $(h^++h^-)/2$ for the model, the model distributions are multiplied by $1/3$. In Figure \ref{fig5} the absolute scale of the experimental particle species dependence is reproduced reasonably well by the model, as well as the ``radial flow effect'' of the slopes decreasing with increasing particle mass, but the absolute slopes from the model at low $m_T-m_0$ are seen to be uniformly somewhat too large (as is also seen in Figure \ref{fig4} at low $p_T$). The model distributions are multiplied by $1/3$ to approximate positive charges.

Although there are clear differences seen in details between model and experiments as described above, it is still remarkable that this simple model does as well as it does in reproducing the overall absolute scales and dependences of these  ``bread and butter'' experimental distributions. It is judged that the description by the model of these ``basic observables'' is adequate enough to cautiously proceed  with using the model to calculate the ``derived observables'' such as elliptic flow, HBT, etc..

\subsection{Elliptic flow}
The elliptic flow variable, $V_2$, is
defined as
\begin{eqnarray}
\label{v2} V_2=<\cos(2\phi)> \\\nonumber
    \phi=\arctan(\frac{p_y}{p_x})
\end{eqnarray}
where ``$<>$'' implies a sum over particles in an event and a sum over events
and where $p_x$ and $p_y$ are the $x$ and $y$ components of the particle
momentum, and $x$ is in the impact parameter direction, i.e.
reaction plane direction, and $y$ is in the direction perpendicular
to the reaction plane. The $V_2$ variable is calculated from the model
using Eq. (\ref{v2}) and taking the reaction plane to be the model $x-z$ plane.

\begin{figure}
\begin{center}
\includegraphics[width=100mm]{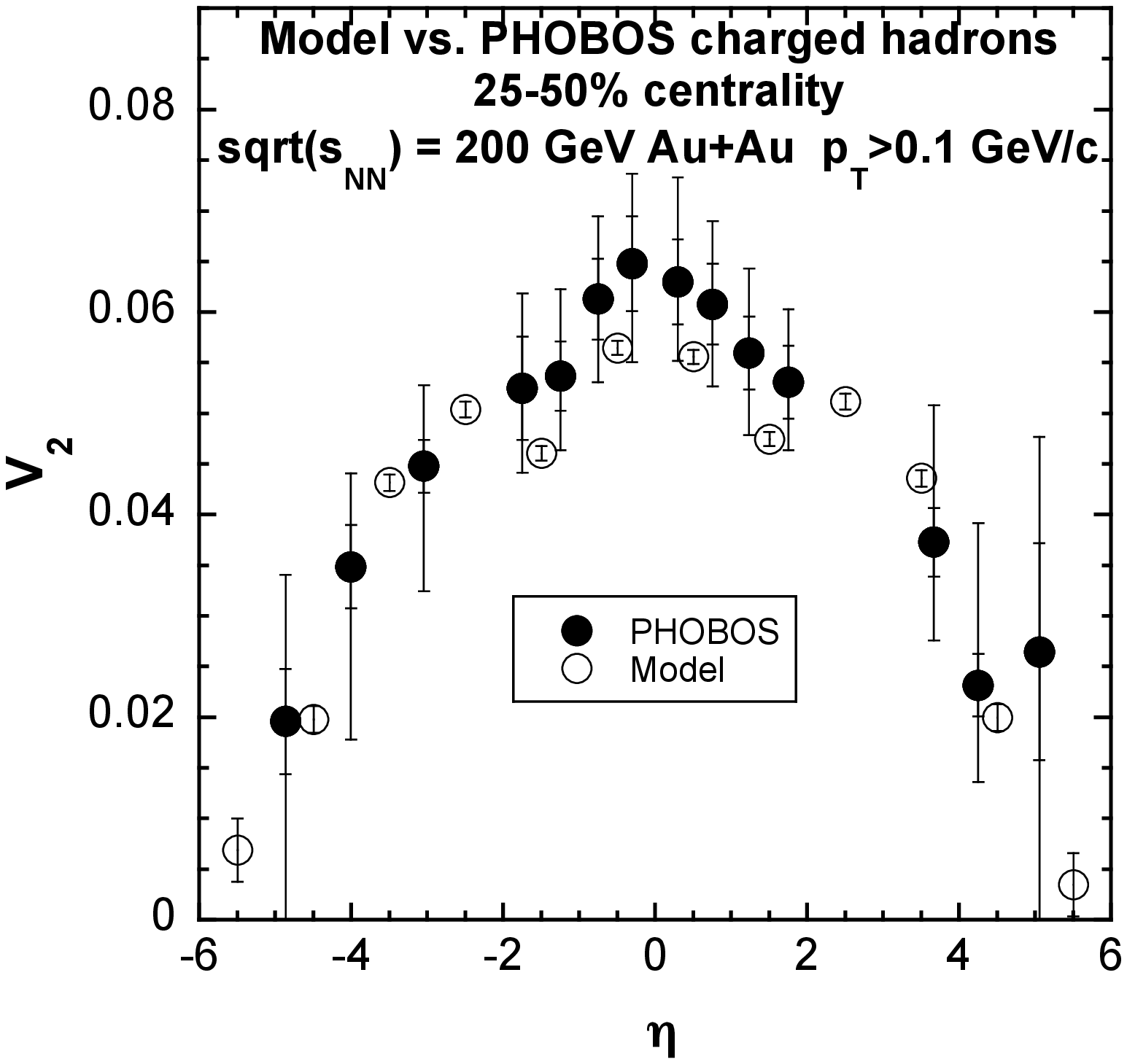} \caption{$V_2$ vs. $\eta$ for Model compared
with PHOBOS 25-50\% centrality. The inner error bars on the PHOBOS data points show the
statistical error and the outer error bars show the statistical+systematic error.}
 \label{fig6}
\end{center}
\end{figure}

\begin{figure}
\begin{center}
\includegraphics[width=100mm]{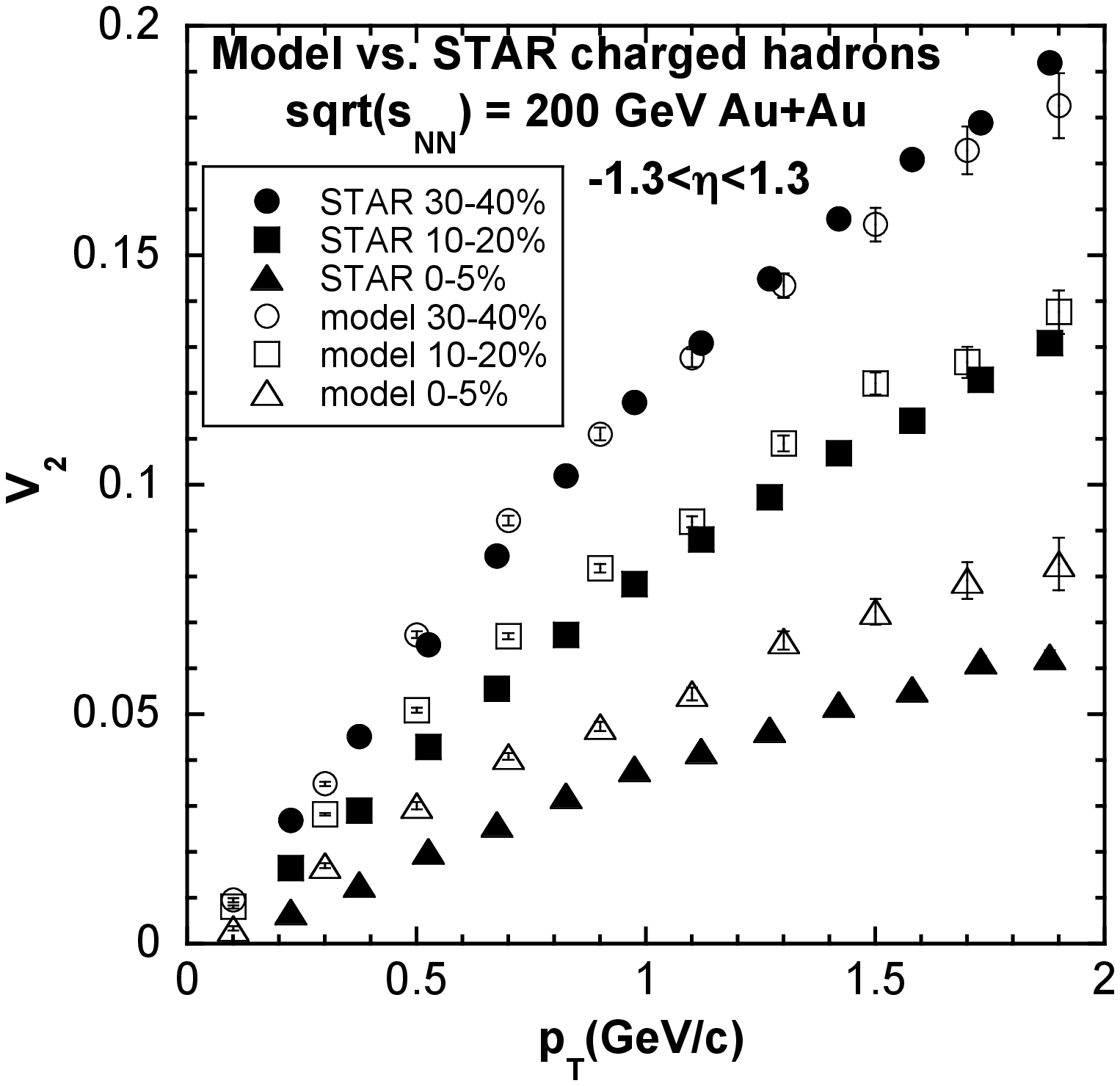} \caption{$V_2$ vs. $p_T$ for Model compared
with STAR for charged particles and several centralities.}
\label{fig7}
\end{center}
\end{figure}

\begin{figure}
\begin{center}
\includegraphics[width=100mm]{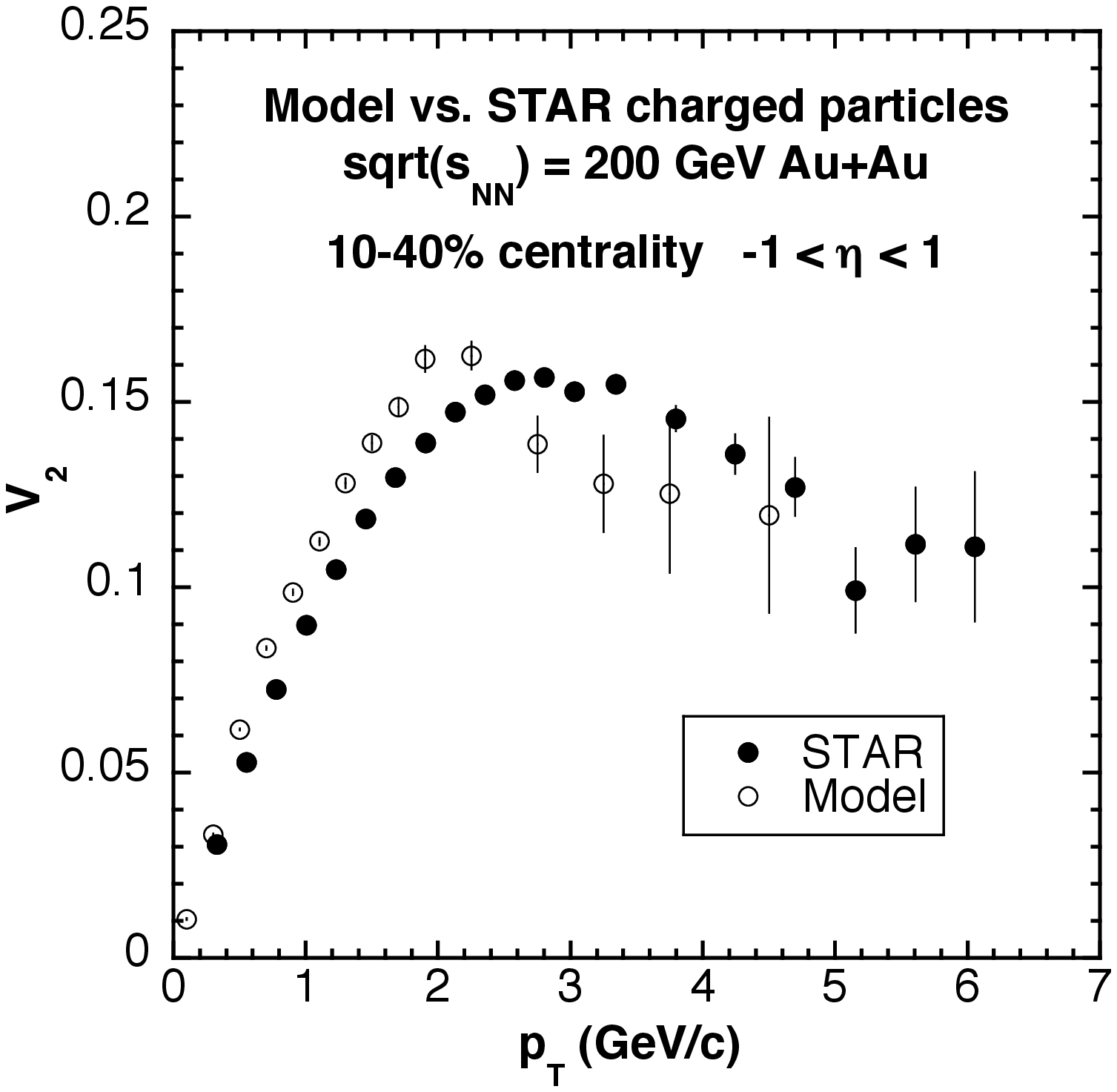} \caption{$V_2$ vs. $p_T$ for Model vs. STAR for
charged particles and up to high $p_T$.}
\label{fig8}
\end{center}
\end{figure}

\begin{figure}
\begin{center}
\includegraphics[width=100mm]{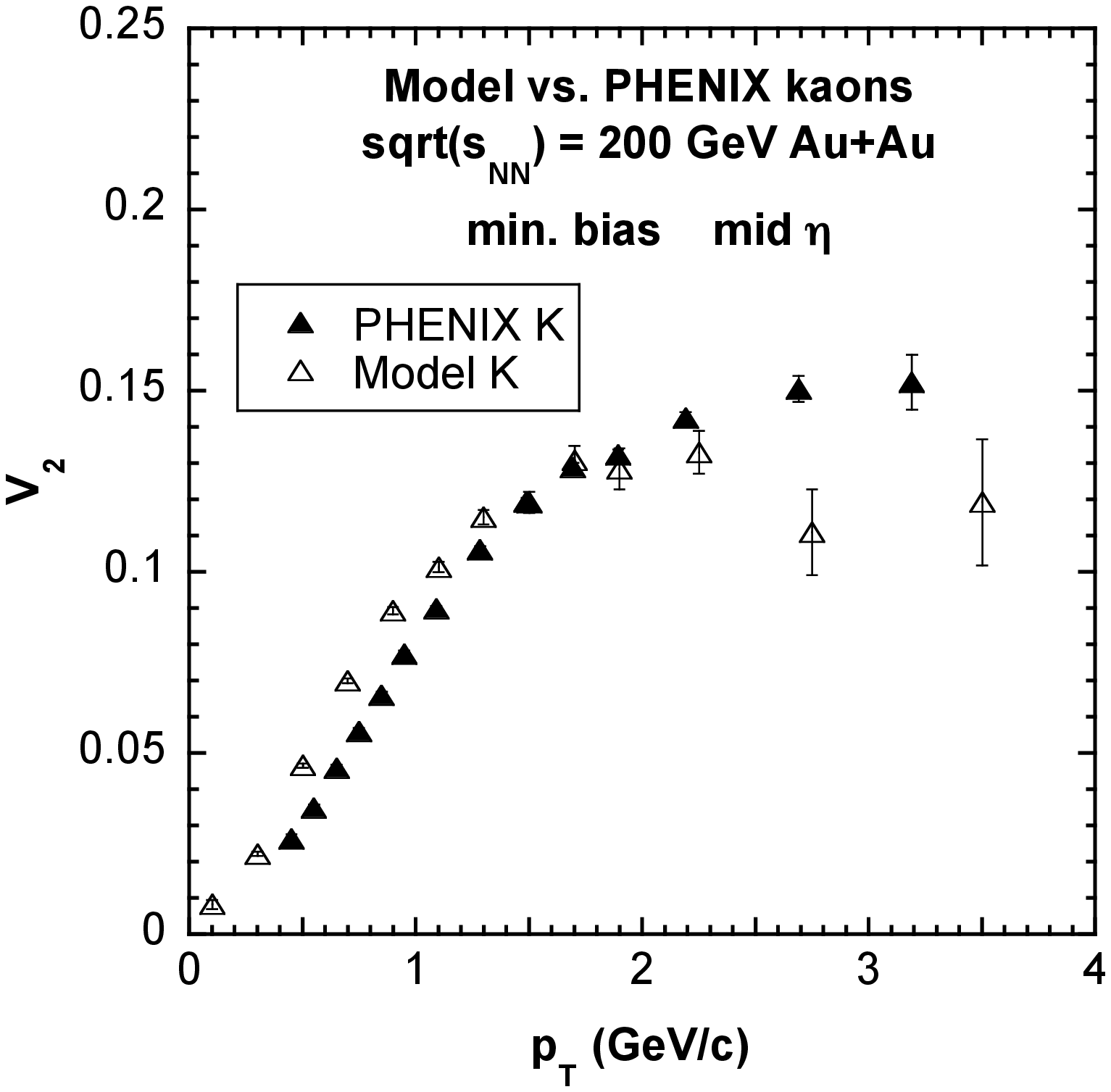} \caption{$V_2$ vs. $p_T$ for Model vs.
PHENIX for kaons. The meaning of ``mid $\eta$''
in this case is that the PHENIX spectra are measured in the range $-0.35<\eta<0.35$, whereas the
model spectra are calculated in the range $-1.3<\eta<1.3$ for better statistics.}
\label{fig9}
\end{center}
\end{figure}

\begin{figure}
\begin{center}
\includegraphics[width=100mm]{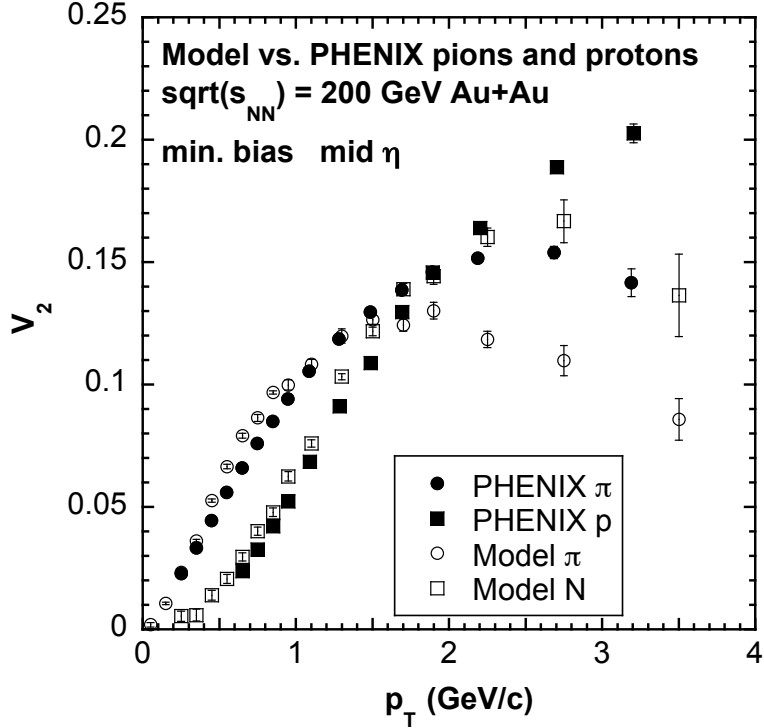} \caption{$V_2$ vs. $p_T$ for Model vs.
PHENIX for pions and protons. The meaning of ``mid $\eta$'' is the same as for Figure \ref{fig9}.}
\label{fig10}
\end{center}
\end{figure}

\begin{figure}
\begin{center}
\includegraphics[width=100mm]{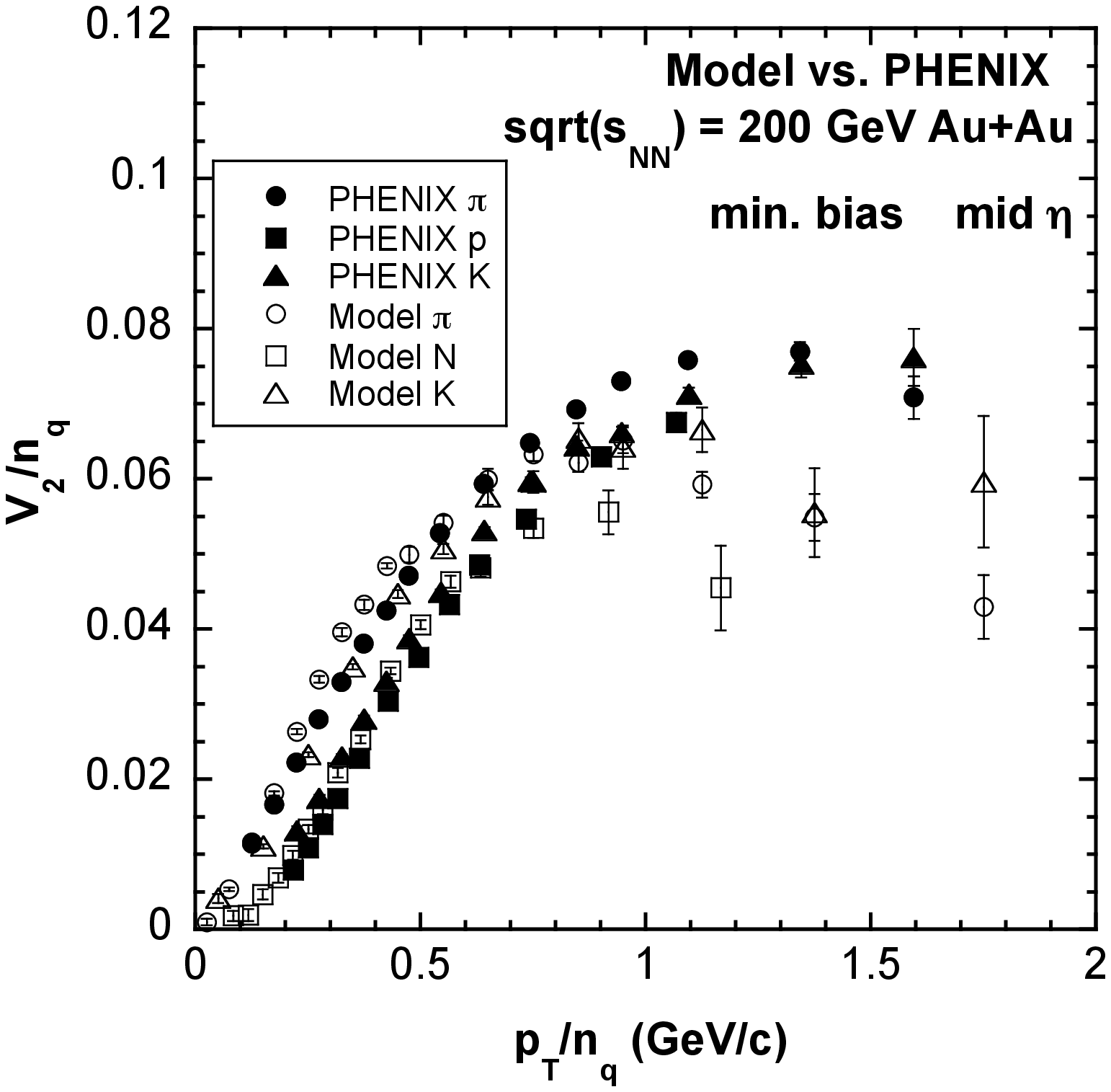} \caption{$V_2/n_q$ vs. $p_T/n_q$ for
Model vs. PHENIX for pions, kaons, and protons.}
\label{fig11}
\end{center}
\end{figure}

Figures \ref{fig6}-\ref{fig11} show comparisons between the model and experiments for elliptic flow.
Figure \ref{fig6} compares the model to PHOBOS for charged particles 
for $V_2$ vs. $\eta$ in a centrality window of
$25-50\%$ \cite{Back:2004mh}. The model is seen to agree with the measurements within error bars
for the entire range in $\eta$, i.e. $-6<\eta<6$, although it looks systematically about 10\% lower than
experiment around mid-rapidity. Note that in the model, $V_2$ is completely determined by
rescattering such that if the rescattering is turned off, $V_2=0$ in all kinematic regions. 

Figure \ref{fig7} shows a comparison of the model to $V_2$ vs. $p_T$ for $p_T<2$ GeV/c for charged particles in several centrality bins from STAR \cite{Adams:2004bi}. The model is seen to
do a reasonable job in representing the different centralities, although it is systematically higher than measurement by about $0.01-0.02$ over the entire $p_T$ range for the $0-5\%$ centrality bin.
Figure \ref{fig8} compares the model with STAR for $V_2$ vs. $p_T$ for 
charged particles in a centrality bin $10-40\%$
in a wide $p_T$ range, i.e. $p_T<6$ GeV/c \cite{Ab:2008ed}. What is 
remarkable about this comparison is that
the model describes the $p_T$ behavior of the experiment in which $V_2$ increases for
$p_T<2$ GeV/c, flattens out, and then starts decreasing for $p_T>3$ GeV/c. Once again, this behavior
is completely rescattering-driven in the model. 

In Figures \ref{fig9} and \ref{fig10} the model is compared with identified-particle $V_2$ vs. $p_T$ plots for kaons, and pions and protons, respectively,
and for minimum bias centrality from
PHENIX \cite{Adare:2006ti}. For both figures the low $p_T$ behavior, i.e. $p_T<2$ GeV/c is described
well and quantitatively by the model, whereas the high $p_T$ behavior, i.e. $p_T>2$ GeV/c is only described qualitatively. In Figure \ref{fig9} the high $p_T$ behavior of the kaons flattens out for both experiment and model, but for the model it flattens out to a slightly lower value. In Figure \ref{fig10} the high $p_T$
behavior for the pions is to flatten out and start decreasing as the model also does but to a lower value
whereas the proton $V_2$ continues increasing, as it does for the model nucleons, but the model does not increase as fast (and the model 3.5 GeV/c point decreases). Figure \ref{fig11} shows the plots in
Figures \ref{fig9} and \ref{fig10} replotted in terms of the number of valence quarks in the
identified particle, $n_q$, as $V_2/n_q$ vs. $p_T/n_q$. The point of doing this is
to show that the different identified particles follow a universal curve when plotted on the same
graph this way. Not surprisingly in the context of the discussion above, the model is seen to follow the experimental scaling quantitatively for
$p_T/n_q<1$  GeV/c and qualitatively at a lower value for $p_T/n_q>1$ GeV/c.

\subsection{Two-pion femtoscopy (Hanbury-Brown-Twiss interferometry)}
For the HBT calculations from the model, the three-dimesional 
two-pion correlation function is formed
and a Gaussian function in momentum difference variables is fitted to it to 
extract the pion source
parameters.  Boson statistics are introduced after the
rescattering has finished (i.e. when all particles have ``frozen out'')
using the standard method of pair-wise symmetrization of bosons in
a plane-wave approximation \cite{Humanic:1986a}. The three-dimensional
correlation function, $C(Q_{side},Q_{out},Q_{long})$, is then calculated 
in terms of the momentum-difference
variables $Q_{side}$, which points in
the direction of the sum of the two pion momenta in the transverse
plane, $Q_{out}$, which points perpendicular to $Q_{side}$ in the
transverse plane and the longitudinal variable along the beam
direction $Q_{long}$.

\begin{figure}
\begin{center}
\includegraphics[width=100mm]{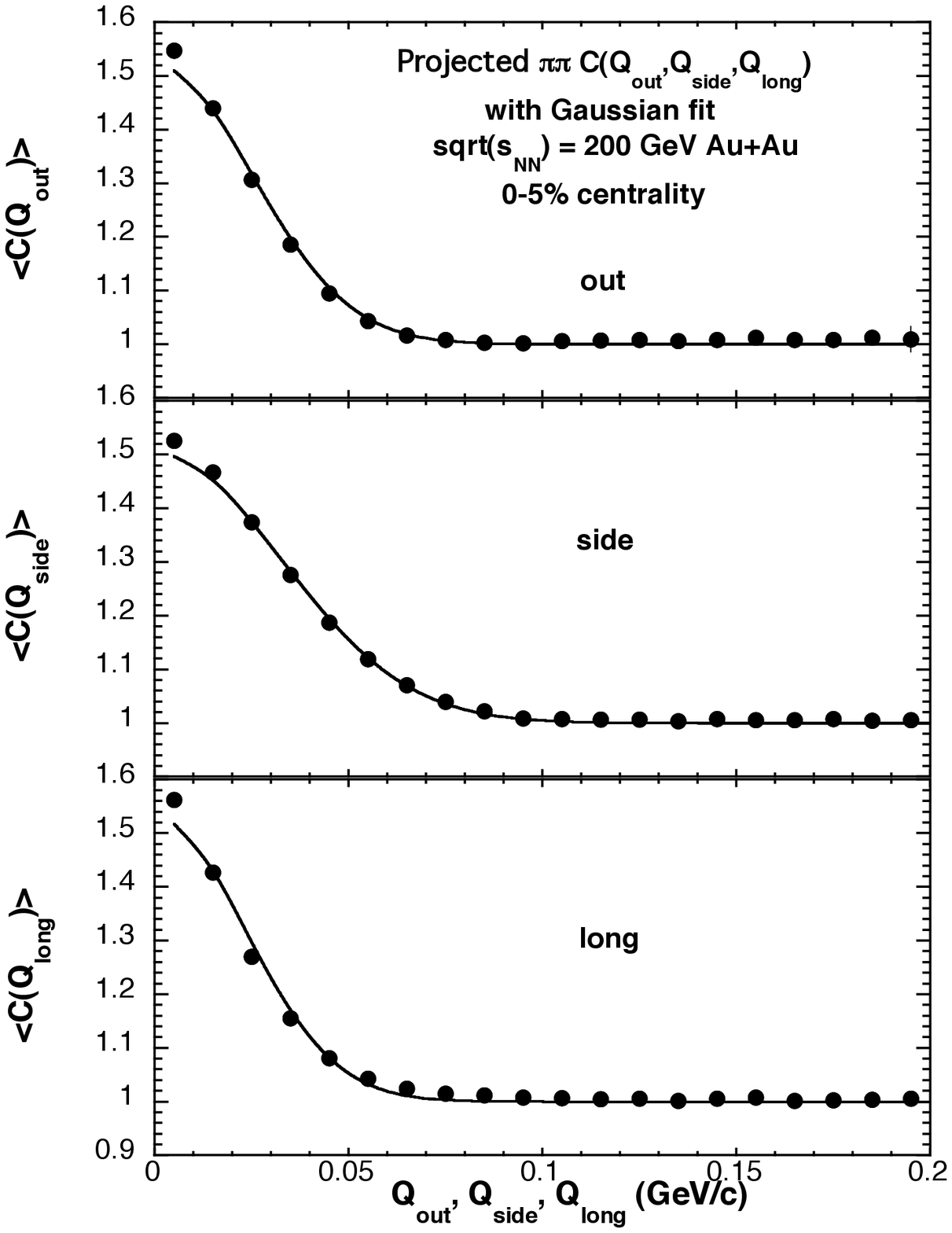} \caption{Sample two-pion correlation
function with Gaussian fit projected onto the $Q_{out}$, $Q_{side}$, and $Q_{long}$
axes from the Model. The collision centrality is $0-5\%$ with cuts on the pions
$-0.5<y<0.5$, $0.15<p_T<0.8$ GeV/c, and $0.15<k_T<0.25$ GeV/c.}
\label{fig12}
\end{center}
\end{figure}

The final step in the calculation is extracting fit parameters by
fitting a Gaussian parameterization to the model-generated two-pion correlation 
function given by, \cite{Lisa:2005a}
\begin{eqnarray}
\label{e6}
\lefteqn{C(Q_{side},Q_{out},Q_{long}) = } \nonumber \\
& & G[ 1 + \lambda \exp( - Q_{side}^{2}R_{side}^{2} -
Q_{out}^{2}R_{out}^{2} - Q_{long}^{2}R_{long}^{2}
-Q_{out}Q_{side}R_{outside}^2) ]
\end{eqnarray}
where the $R$-parameters, called the radius parameters, are associated with each 
momentum-difference variable direction, G is a normalization constant, and
$\lambda$ is the usual empirical parameter added to help in the
fitting of Eq. (\ref{e6}) to the actual correlation function
($\lambda = 1$ in the ideal case). The fit is carried out in the conventional LCMS 
frame (longitudinally comoving system) in which the
longitudinal boson pair momentum vanishes \cite{Lisa:2005a}.
Figure \ref{fig12} shows a sample projected correlation function from the model with projected fit
to Eq. (\ref{e6}).

\begin{figure}
\begin{center}
\includegraphics[width=150mm]{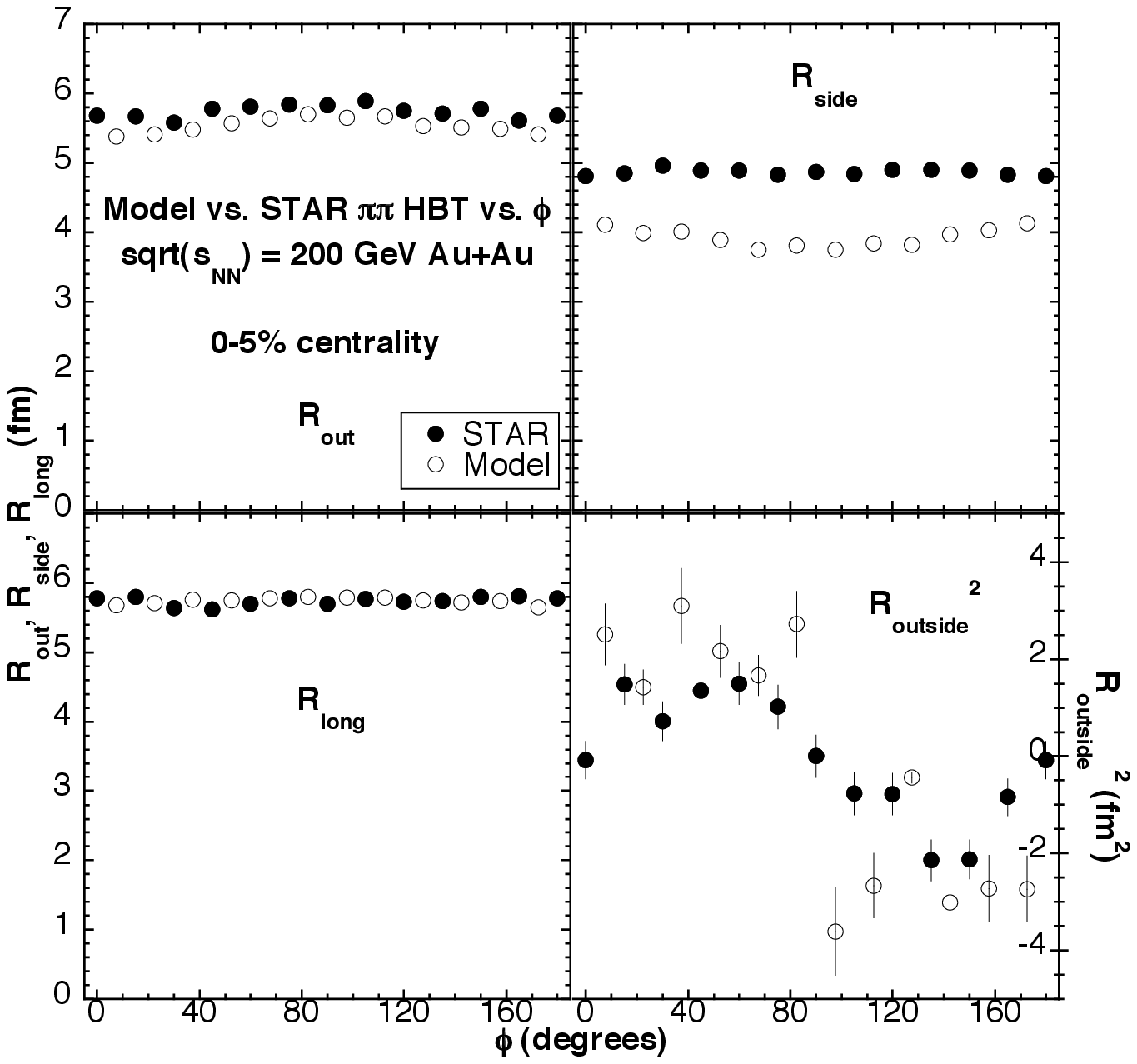} \caption{Azimuthal two-pion HBT 
parameters vs. $\phi$ from Model vs. STAR for centrality 0-5\%. Pions are accepted in
the cut ranges $-1<y<1$, $0.1<p_T<0.6$ GeV/c, and $0.15<k_T<0.6$ GeV/c.}
\label{fig13}
\end{center}

\end{figure}

\begin{figure}
\begin{center}
\includegraphics[width=150mm]{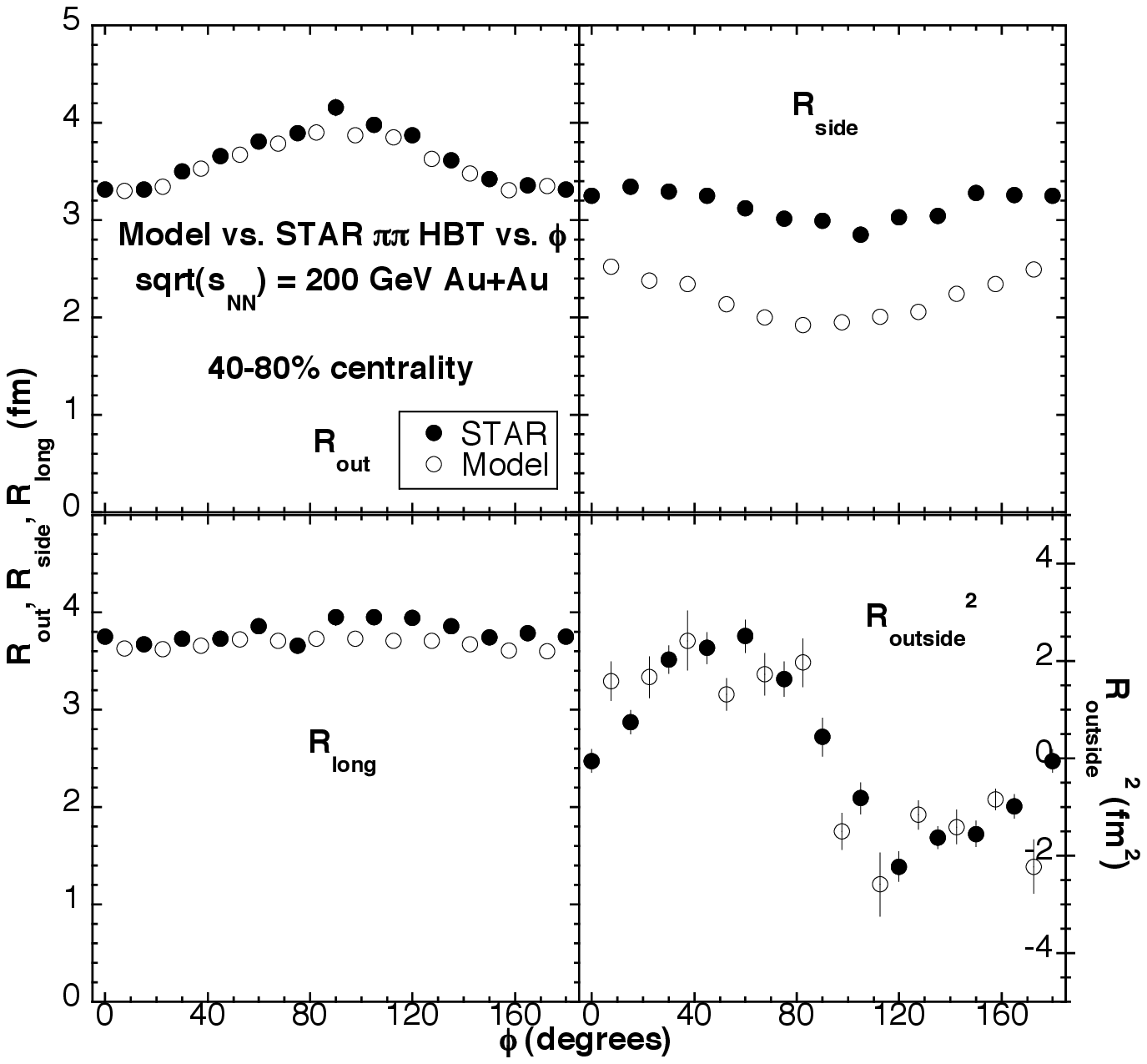} \caption{Azimuthal two-pion HBT 
parameters vs. $\phi$ from Model vs. STAR for centrality 40-80\%. Pions are accepted in
the cut ranges $-1<y<1$, $0.1<p_T<0.6$ GeV/c, and $0.15<k_T<0.6$ GeV/c.}
\label{fig14}
\end{center}
\end{figure}

Figures \ref{fig13} and \ref{fig14} show comparisons between the model and STAR
for radius parameters extracted as a function of azimuthal angle, $\phi$, for two centrality
cuts, $0-5\%$ and $40-80\%$, respectively \cite{Adams:2003ra}. In Figure \ref{fig13} the model
is seen to describe the more or less "flat" dependence on $\phi$ 
for $R_{out}$, $R_{side}$, and $R_{long}$
seen in the experiment for these central collisions, although the model is seen to under predict
the magnitude of $R_{side}$ by about 20\%. The model is seen 
to follow the trend of the experiment for
$R_{outside}^2$ within the large statistical error bars shown. For the less central collisions
shown in Figure \ref{fig14}, the model describes the oscillatory behavior now seen in $R_{out}$
and $R_{side}$ as well as $R_{outside}^2$ and the continued flat dependence seen in
$R_{long}$, although once again under predicting the magnitude of $R_{side}$ by about
30\%. The $\lambda$-parameters extracted in the fits from the model were constant
in $\phi$ with the values 0.61 and 0.54 for the $0-5\%$ and $40-80\%$ centrality bins,
respectively.


\begin{figure}
\begin{center}
\includegraphics[width=150mm]{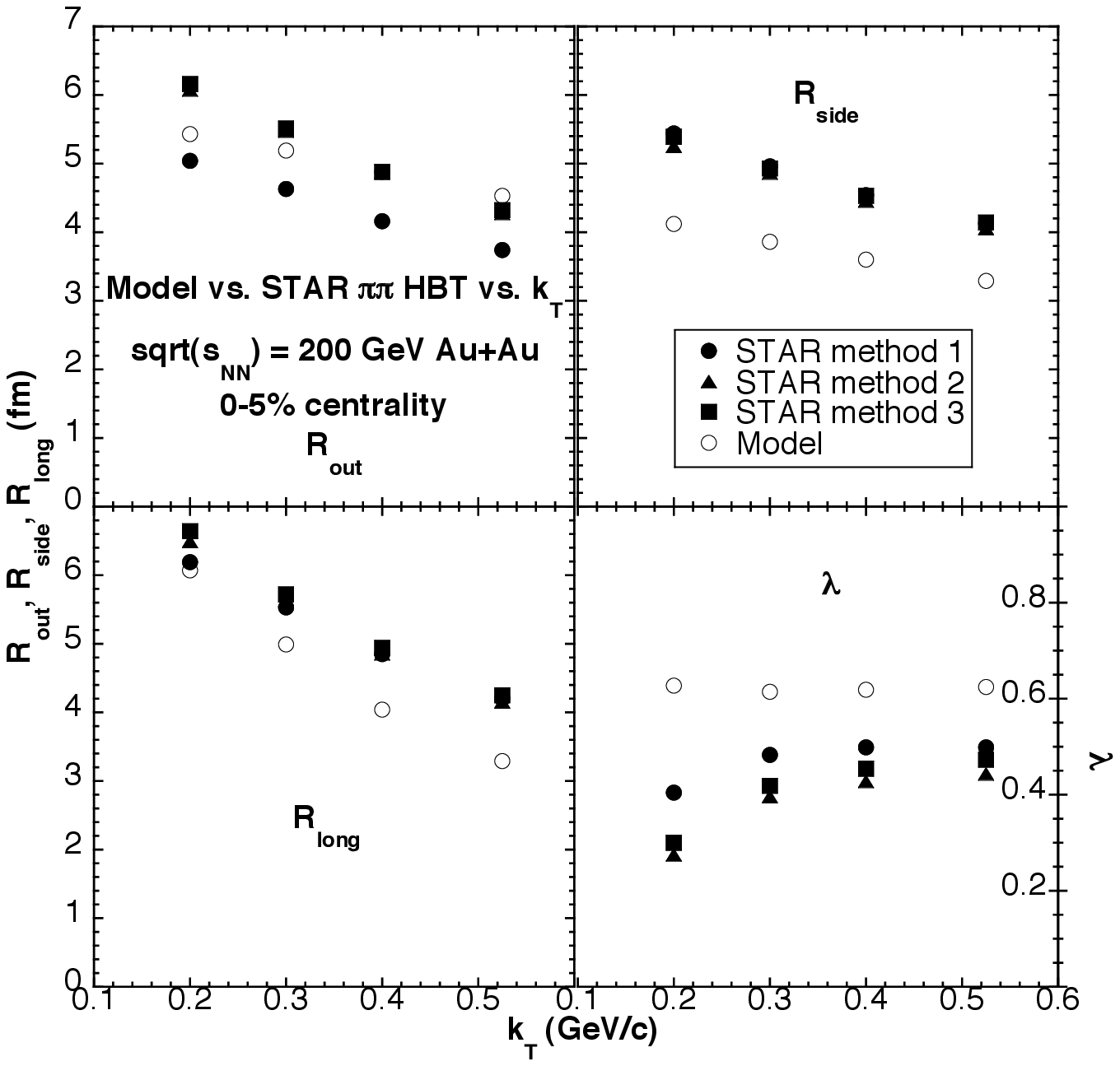} \caption{Two-pion HBT 
parameters vs. $k_T$ from Model vs. STAR for centrality 0-5\%. Pions are accepted in
the cut ranges $-0.5<y<0.5$ and $0.15<p_T<0.8$ GeV/c. The meaning of the
three methods used by STAR to extract parameters may be found in Ref. \cite{Adams:2004yc}}
\label{fig15}
\end{center}
\end{figure}

Figure \ref{fig15} compares the model with STAR for the $k_T$ dependence of the radius
parameters in a centrality bin of $0-5\%$ \cite{Adams:2004yc}. For these fits, the parameter
$R_{outside}$ in Eq. (\ref{e6}) is set to zero. Several methods used by STAR in extracting
their fit parameters were used and as seen they give approximately the same basic
 results \cite{Adams:2004yc}. The model is seen to follow the general trend of the experiment for
 decreasing radius parameters for increasing $k_T$. The model slightly over predicts this effect
 for $R_{long}$ and, as was the case in the azimuthal HBT results
 shown earlier,  consistently under predicts the magnitude of 
 $R_{side}$ by about 24\%. As seen in the experiment, the model $\lambda$-parameter
 is less than 1, being constant in $k_T$ at about 0.61, but larger than the experimental values on
 average by about 30\%. Note that the main source of  $\lambda<1$ in the model is from the
 presence of long-lived resonances such as $\eta$ and ${\eta}'$.

\subsection{High $p_T$}
Studying the high $p_T$ behavior of the observables $R_{AA}$ and $dn/d\Delta\phi$
is thought to be a way of more directly studying QCD processes, such as jets, 
in heavy-ion collisions. As mentioned earlier, since the present model is based on
using PYTHIA which uses QCD processes in calculating $p+p$ collisions, the model
should contain these effects and thus should be suitable for comparing with experiments
which measure these observables.

The $R_{AA}$ is defined as \cite{Adler:2003au},
\begin{eqnarray}
\label{e7}
R_{AA}=(\frac{1}{N_{ev}}\frac{d^2N^{AuAu}}{dp_Td\eta})/(T_{AuAu}\frac{d^2\sigma^{pp}}{dp_Td\eta})
\end{eqnarray}
where the numerator is the usual $Au+Au$ $p_T$ distribution as shown in Figure \ref{fig4} and the
denominator has the $p+p$ $p_T$ distribution normalized to cross section as in Figure \ref{fig1}
and multiplied by the quantity $T_{AuAu}$, the Glauber nuclear overlap function which is
different for each centrality cut.

\begin{figure}
\begin{center}
\includegraphics[width=140mm]{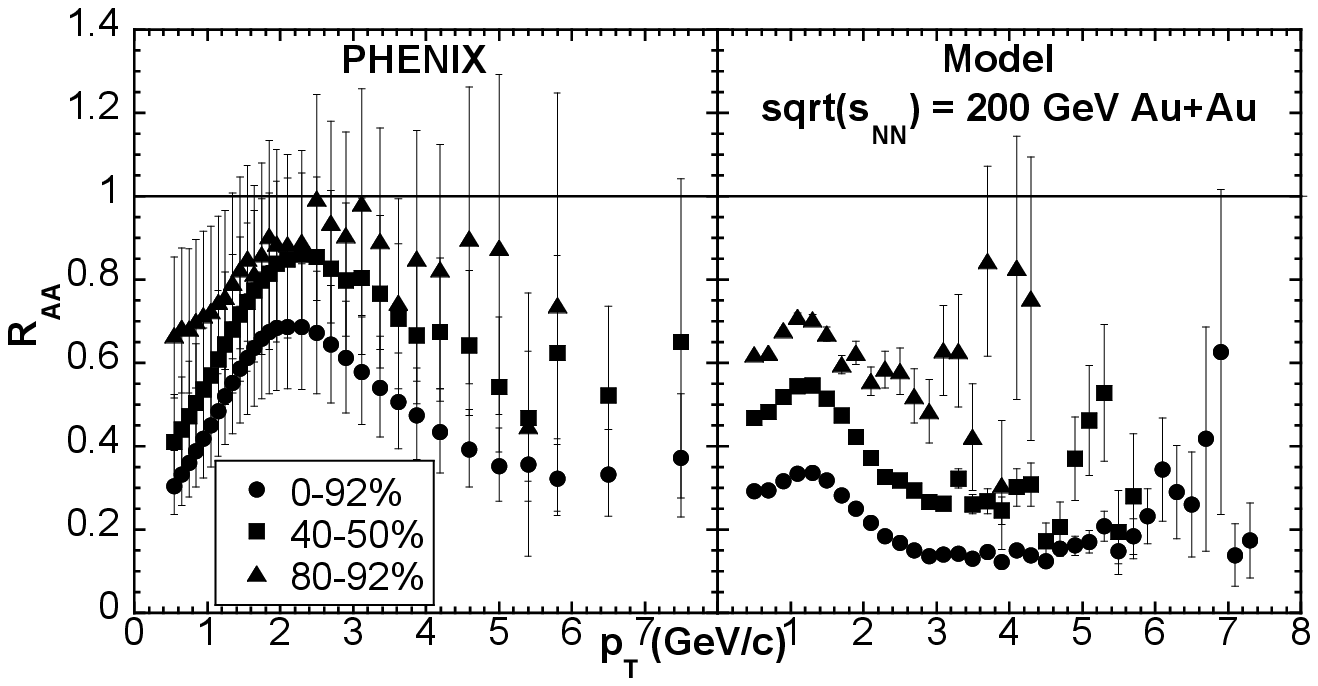} \caption{$R_{AA}$ for Model compared
with PHENIX for several centralities.}
\label{fig16}
\end{center}
\end{figure}

Figure \ref{fig16} compares the model to PHENIX for $R_{AA}$ vs. $p_T$ for 
three centrality windows \cite{Adler:2003au}. For both the model and PHENIX 
the plots in Figures \ref{fig1}
and \ref{fig4} were used in Eq. (\ref{e7}) to calculate these $R_{AA}$ plots along with
the $T_{AA}$ values shown in Table I. of Ref. \cite{Adler:2003au}. The error bars shown
for the PHENIX plots are a sum of both statistical error and the overall scale uncertainty, 
and they mostly
reflect the scale uncertainty. As seen, the model describes three main 
qualitative features of the experiment:
1) for large $p_T$ the $R_{AA}$ decreases with increasing $p_T$, and as the centrality window
goes from minimum bias ($0-92\%$) to peripheral ($80-92\%$) 2)  the scale of $R_{AA}$
increases and 3) the dependence of $R_{AA}$
on $p_T$ tends to flatten out. It is also seen that, even with the uncertainty in the PHENIX
overall normalization, the model scale tends to be lower than experiment, and at low $p_T$
the peaks in the plots for the model occurs at about 1.3 GeV/c whereas the peaks occur
at about 2.3 GeV/c for experiment. These differences of the model with experiment reflect 
the differences already seen in the $p_T$ distributions in Figure \ref{fig4}, but the qualitative
similarities as discussed above are clearly present out to the highest $p_T$ shown.

\begin{figure}
\begin{center}
\includegraphics[width=120mm]{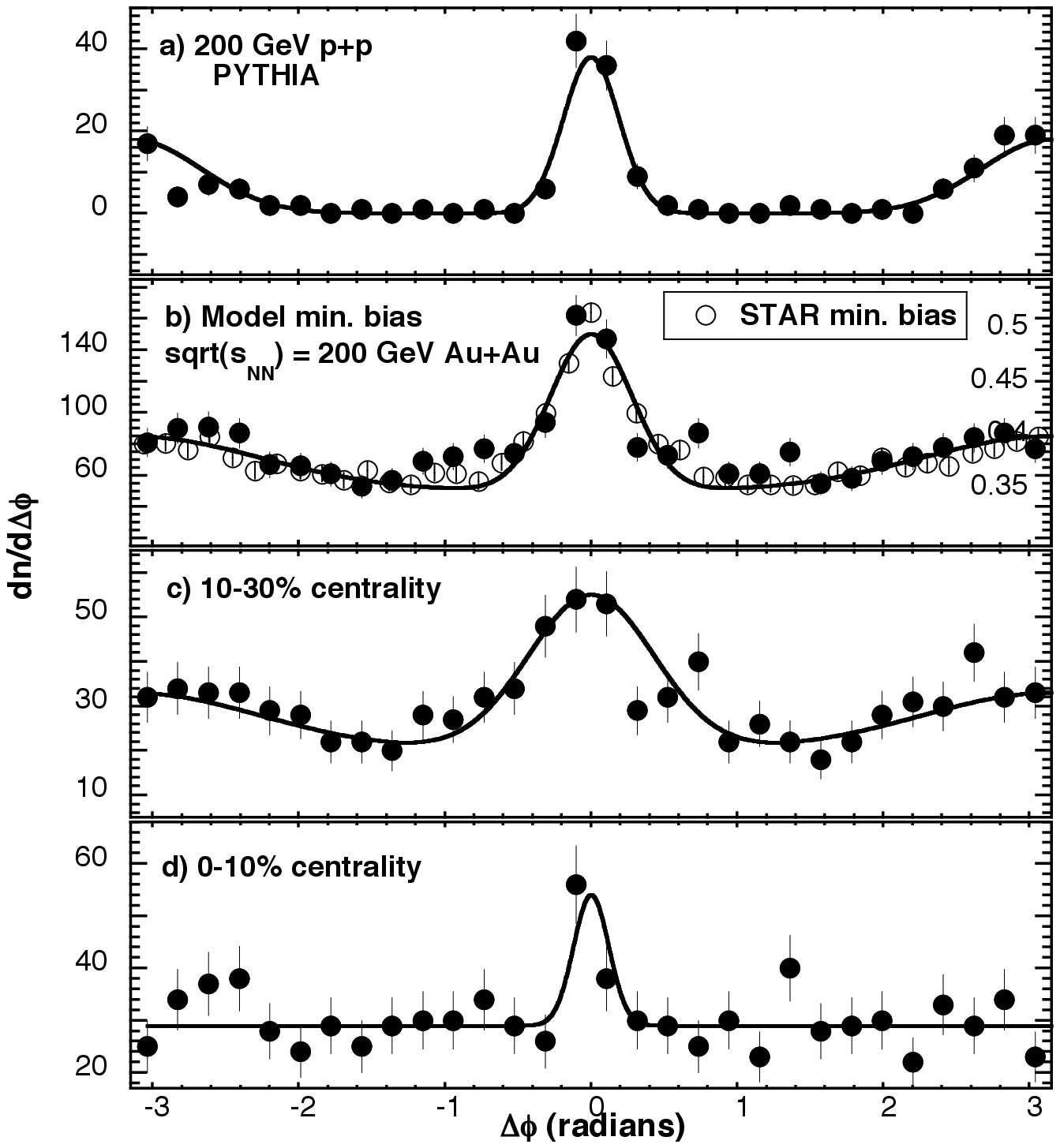} \caption{$dn/d\Delta\phi$ vs. $\Delta\phi$ plots using STAR
cuts on $p_T$. a) 200 GeV $p+p$ collisions from PYTHIA. 
b) $\sqrt{s_{NN}}=200$ GeV $Au+Au$ with minimum
bias centrality for Model and STAR. 
c) $\sqrt{s_{NN}}=200$ GeV $Au+Au$ for Model with $10-30\%$ centrality. 
d) Same as c) with $0-10\%$ centrality. Fits to the points are also shown to guide the eye.}
\label{fig17}
\end{center}
\end{figure}

Figure \ref{fig17} shows $dn/d\Delta\phi$ vs. $\phi$ plots from the model and a comparison
of one of them with STAR charged particles \cite{Adler:2002tq}. The model plots, which include all
hadrons, are made using the same cuts on rapidity and $p_T$ as used by STAR, namely
for individual particles $\mid\eta\mid<0.7$ and $p_T> 2$ GeV/c, and for particle pairs,
one of which is a ``trigger particle'', from which $\Delta\phi$ is formed, $\mid\Delta\eta\mid<1.4$ and $p_T^{Trig}>4$ GeV/c. The lines are fits to the model points to guide the eye and the model
$dn/d\Delta\phi$ normalizations are in counts per bin. Figure \ref{fig17}a
shows a plot from PYTHIA for 200 GeV $p+p$ as a reference. The forward and backward peaks
from di-jet production are clearly seen at $\Delta\phi=0$ and $\pm\pi$, respectively. Figure \ref{fig17}b
compares the model to STAR minimum bias 
$\sqrt{s_{NN}}=200$ GeV $Au+Au$ collisions. The scale for the STAR
plot is shown on the right-hand axis of the figure. As seen the model describes the shape of
the experiment well, both the width of the forward peak and the relative height of the
backward peak with respect to the height of the forward peak. Figure \ref{fig17}c shows a
model plot for a medium centrality window at $10-30\%$. Features similar to the minimum bias
plot are seen with a forward and backward peak of similar relative heights, although the width
of the forward peak is somewhat larger than in the minimum bias case. A more central case
from the model is shown in Figure \ref{fig17}d where the centrality window is $0-10\%$. Although
this plot is pushing the edge of the statistics possible from the 87K event model run used in this
study for $\sqrt{s_{NN}}=200$ GeV $Au+Au$ collisions, it appears to have a qualitatively different shape
compared with the other plots shown in this figure. Namely, besides the presence of the forward
peak, the plot looks more or less flat for values of $\Delta\phi$ out to $\pm\pi$, i.e. the backward peak
appears suppressed. This is the same general behavior seen in STAR in the same centrality
window, i.e. Figure 1c of Ref. \cite{Adler:2002tq}.

\section{Predictions from the model for $Pb+Pb$ collisions at $\sqrt {s_{NN}} = 5.5$ TeV}
Predictions from the model for LHC-energy collisions for $\sqrt {s_{NN}} = 5.5$ TeV
$Pb+Pb$ collisions are presented below. These will be sample predictions for $p_T$ and $\eta$
distributions, elliptic flow and HBT to give a flavor of the differences predicted by
the model for LHC $Pb+Pb$ collisions compared with RHIC $Au+Au$ collisions. In this spirit,
RHIC-energy $Au+Au$ collisions with the same kinematic conditions as the LHC-energy
$Pb+Pb$ collisions will be shown to indicate the trends of the predictions.
With the changes described above in Section II, the model will be used in the same
way to make the $\sqrt{s_{NN}}=5.5$ TeV $Pb+Pb$
predictions as it was used for the $\sqrt{s_{NN}}=200$ GeV $Au+Au$
calculations, including the use of the short proper hadronization time of $\tau=0.1$ fm/c.

\begin{figure}
\begin{center}
\includegraphics[width=100mm]{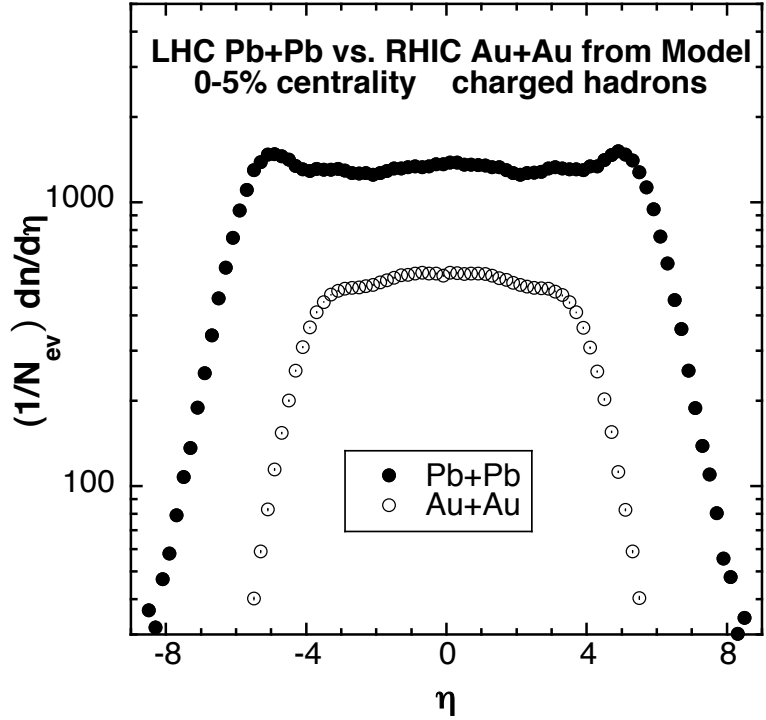} \caption{Rapidity distributions from Model
comparing $Pb+Pb$ collisions at $\sqrt {s_{NN}} = 5.5$ TeV (LHC) with $Au+Au$ collisions at 
$\sqrt {s_{NN}} = 200$ GeV (RHIC) for charged particles and $0-5\%$ centrality.}
\label{fig18}
\end{center}
\end{figure}

\begin{figure}
\begin{center}
\includegraphics[width=100mm]{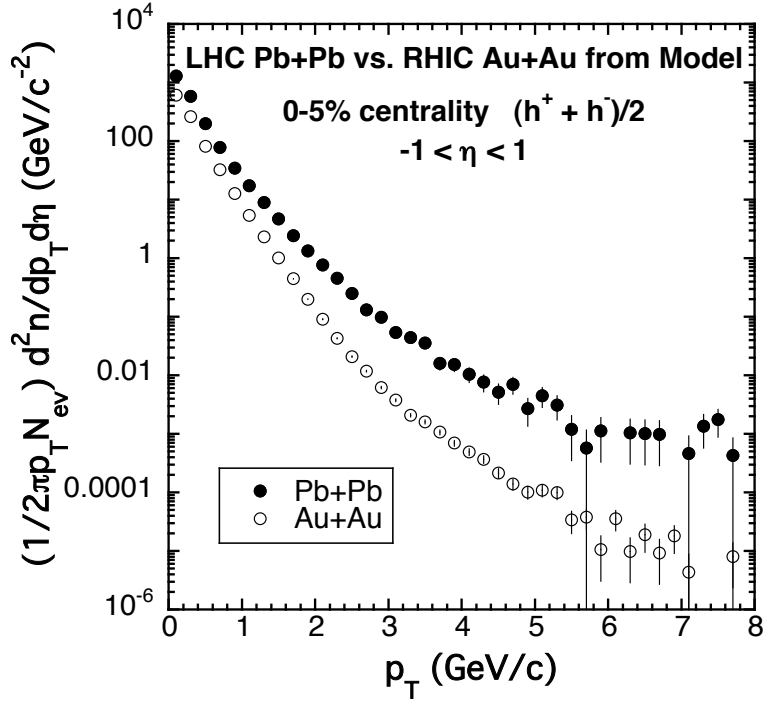} \caption{$p_T$ distributions from Model
comparing LHC $Pb+Pb$ and RHIC $Au+Au$ collisions.}
\label{fig19}
\end{center}
\end{figure}

\begin{figure}
\begin{center}
\includegraphics[width=100mm]{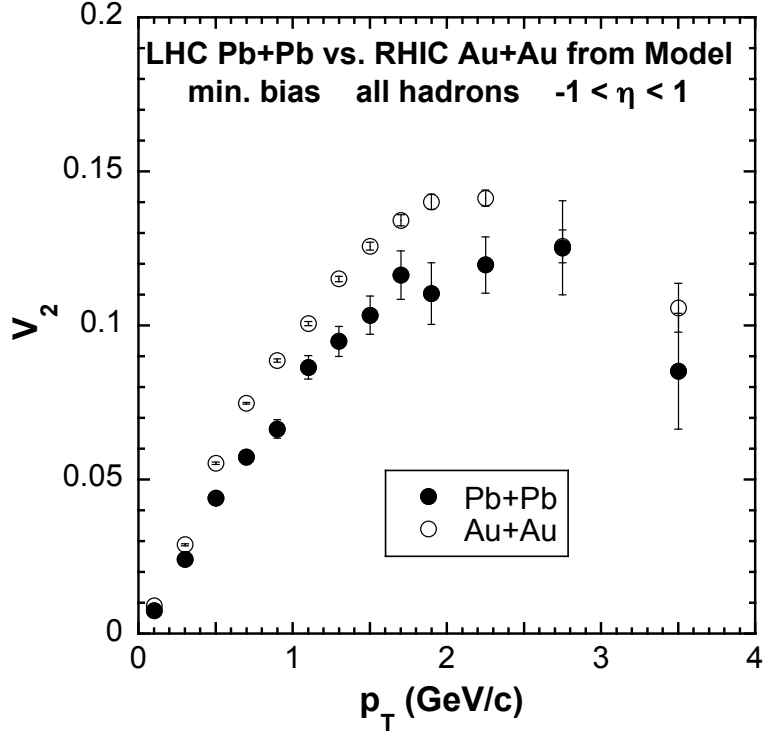} \caption{$V_2$ vs. $p_T$ from Model comparing
LHC $Pb+Pb$ with RHIC $Au+Au$ collisions;
minimum bias centrality, all hadrons, and $-1<\eta<1$.}
\label{fig20}
\end{center}
\end{figure}

\begin{figure}
\begin{center}
\includegraphics[width=140mm]{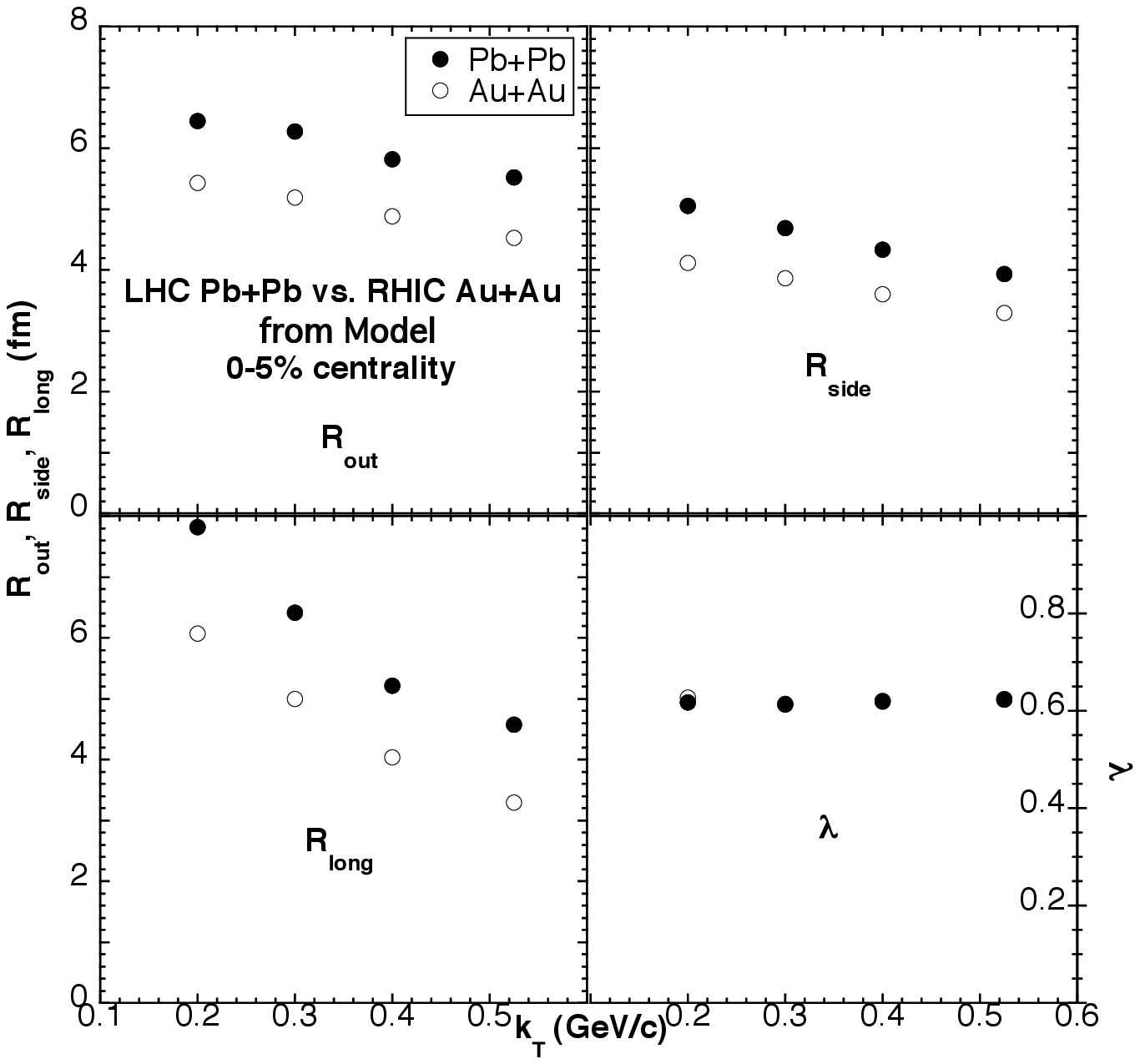} \caption{Two-pion HBT vs. $k_T$ from Model comparing
LHC $Pb+Pb$ with RHIC $Au+Au$ collisions; $0-5\%$ centrality,
$-0.5<y<0.5$, and $0.15<p_T<0.8$ GeV/c.}
\label{fig21}
\end{center}
\end{figure}

Figures \ref{fig18} - \ref{fig21} show the model predictions for 
$\sqrt{s_{NN}}=5.5$ TeV $Pb+Pb$ collisions
compared with $\sqrt{s_{NN}}=200$ GeV $Au+Au$ collisions, also from the model.
The most noticeable features
of these predictions are summarized below:
\begin{itemize}
\item $dn/d\eta$ near mid-rapidity for charged particles is predicted to be about 1400 for
a $0-5\%$ centrality window at the LHC. -- This is seen in Figure \ref{fig18}, which predicts that
the mid-rapidity charged particle density for LHC $Pb+Pb$ will be about a factor of 2.5 greater than for RHIC $Au+Au$. From Figure \ref{fig3} it was seen that the model density was about 10\% lower
than experiment. Thus even boosting the LHC prediction up by 10\% puts its value at the lower end of the range of predictions which have been recently made of $1500-4000$ in central collisions using various extrapolations of RHIC experimental rapidity densities \cite{Aamodt:2008zz}.
\item The charged particle $p_T$ distribution for $p_T>5$ GeV/c is predicted to be about two orders of magnitude larger at the LHC compared with RHIC. -- This is seen in Figure \ref{fig19} and is
an expected consequence of the higher $\sqrt {s_{NN}}$ in the 
LHC collisions that the $p_T$ distributions
at high $p_T$ should be greatly enhanced. 
\item Elliptic flow in minimum bias centrality collisions is predicted to be slightly smaller at
the LHC compared with RHIC. -- As seen in Figure \ref{fig20} the plot of $V_2$ vs. $p_T$
for all hadrons for LHC $Pb+Pb$ looks similar to that for RHIC $Au+Au$ collisions, but overall the LHC plot gives a slightly smaller $V_2$ for the entire range in $p_T$. This is somewhat unexpected
since in this model $V_2$ is produced exclusively by the rescattering process, and since the
rapidity density seen in Figure \ref{fig18} is higher at LHC, it might be expected that the more rescattering would result in a higher $V_2$ than for RHIC. Rather than this, it seems as though the
initial ``almond-shaped'' geometry present in the $b>0$ collisions starts to become ``washed
out'' from the enhanced number of rescatterings. This will be discussed more below.
\item $\pi\pi$ HBT radius parameters in $0-5\%$ centrality collisions are predicted to be
$20-30\%$ larger at the LHC compared with RHIC. -- In Figure \ref{fig21} the behavior of the
$k_T$ dependence of the HBT parameters for LHC $Pb+Pb$ is similar to what is seen for
RHIC $Au+Au$, but the overall scale for the radius parameters is slightly larger. 
\end{itemize}

\section{Discussion}
Although some discussion of the individual results has been given above, it is useful to reflect
on the sources of the effects in the present model which produce these
results and to describe what interconnections may exist among them. Whereas PYTHIA provides
the baseline $p+p$ kinematics, the hadronic rescattering is responsible for all collective effects
in the model beyond $p+p$. As seen in Figure \ref{fig2}, the rescattering rate is large in the
early stage of the collision with some rescattering persisting to times of 50 fm/c and beyond.
The radial flow effects seen in the identified particle $m_T$ distributions of Figure \ref{fig5}
and $k_T$ dependence of the HBT parameters in Figure \ref{fig15} as well as the elliptic flow
effects seen in the $V_2$ plots of Figures \ref{fig6}-\ref{fig11} are mostly established in the early
stage, i.e. by $t \approx 10$ fm/c (see Figure 18 of Reference \cite{Humanic:2006a}
for a calculation of $\sqrt {s_{NN}}=200$ GeV Au+Au collisions using
a model similar to the present one). Comparison of the model with the experimental
azimuthal HBT results shown in Figures \ref{fig13} and \ref{fig14} provides a double test
of the flow generated by the rescattering since both radial flow effects, i.e. the sizes of the
HBT parameters,  and elliptic flow effects, i.e. their $\phi$ dependence,  are
present. While the present model gives a reasonably good description of the
experimental elliptic flow results using rescattering effects alone, other cascade
studies have found that either it is not possible to generate large enough elliptic flow \cite{Bleicher:1999xi,Li:2007yd} or it is necessary to use extreme elastic parton
cross sections \cite{Molnar:2001ux} to describe experiments. Although there are many differences
in detail between the present study and those studies, the basic feature allowing the present model to
generate enough elliptic flow is that hadronic rescattering is allowed to take place from the earliest
times and thus during the highest densities, as seen in Figure \ref{fig2}. A concern regarding
carrying out transport calculations at high densities is that non-physical superluminal artifacts can be
introduced which can effect the results \cite{Molnar:2001ux}. 
As mentioned earlier, a study of this effect has been
carried out for a model very similar to the present one and it was found that the observables from the
model studied, i.e. spectra, elliptic flow, and HBT, were not 
significantly affected \cite{Humanic:2006b}. Thus it is considered unlikely that these
artifacts play a significant role in the present study.

This last statement should also be true for the $\sqrt {s_{NN}}=5.5$ Tev $Pb+Pb$
predictions shown in Figures \ref{fig18}-\ref{fig21}. Although the rapidity density at
mid-rapidity for
LHC $Pb+Pb$ is seen to be more than twice as large as for RHIC $Au+Au$ in Figure \ref{fig18},
the mid-rapidity particle density for LHC $Pb+Pb$ for early times is found to be similar to that
for RHIC  $Au+Au$ seen in Figure \ref{fig2}. This is due to the larger hadronization volume
and time resulting from the higher average particle momenta generated in the LHC-energy
collisions, as seen in Figure \ref{fig19}, and calculated from Eqs. (1)-(4). This similarity
between the particle densities would explain why the flow effects seen in 
Figures \ref{fig20} and \ref{fig21} are similar, i.e the similarity between the LHC and RHIC
elliptic flow and the dependence of the HBT radius parameters on $k_T$ (although the
overall size of the LHC radius parameters is greater).

The features of the high $p_T$ observables $R_{AA}$ and $dn/d\Delta\phi$ calculated
from the model and shown in Figures \ref{fig16} and \ref{fig17} are also driven by the
underlying rescattering. For these observables it is the energy loss of the $p+p$ produced
high $p_T$ particles rescattering 
with the rest of the particles that creates the effects. The effect of the rescattering can be
seen for 
$R_{AA}$  by considering that if the rescattering 
were turned off in the model the numerator in Eq. \ref{e7}, which
would be simply a superposition of $p+p$ events,
would be exactly the same distribution as the denominator except for a scale factor, resulting
in a flat dependence on $p_T$ for all centralities. The same would be the case for $dn/d\Delta\phi$
in Figure \ref{fig17} if rescattering were turned off: Figure \ref{fig17}d would look like
Figure \ref{fig17}a.

\section{Summary and Conclusions}
A simple kinematic model based on superposition of $p+p$ collisions, 
relativistic geometry and final-state hadronic rescattering
has been used to calculate various hadronic observables in $\sqrt {s_{NN}} = 200$ GeV
$Au+Au$ collisions and $\sqrt {s_{NN}} = 5.5$ TeV $Pb+Pb$ collisions. The model calculations
were compared with experimental results from several $\sqrt {s_{NN}} = 200$ GeV $Au+Au$ 
collision studies from RHIC. With the short hadronization time assumed in the model of $\tau=0.1$ fm/c, it is found that this model 
describes the trends of the observables from these experiments
surprisingly well considering the model's simplicity. This also gives more credibility to the
model predictions presented for LHC-energy $\sqrt {s_{NN}} = 5.5$ TeV $Pb+Pb$ collisions. 

As shown above, the main strength of the present model is not that it gives precise agreement with experiment for individual observables in particular kinematic regions, but in its ability to give an overall qualitative description of a range of observables in a wide kinematic region, i.e. to summarize the gross features seen in experiments for $\sqrt {s_{NN}} = 200$ GeV $Au+Au$ collisions. Another strength is its
simplicity. Besides the kinematics generated in the superposed $p+p$ collisions by PYTHIA, the only
other ``active ingredient'' in the model driving the kinematics underlying the hadronic observables shown is the final-state hadronic rescattering. As discussed above,
if the hadronic rescattering were turned off in the model,
all elliptic flow would disappear, the HBT radius parameters would lose all $\phi$ and $k_T$ dependence and would be significantly smaller, the $R_{AA}$ vs. $p_T$ plots would be flat, and
all of the $dn/d\Delta\phi$ plots would look like $p+p$ (i.e. Figure \ref{fig17}a). 
Making $\tau$ large is another way to effectively turn off all of these effects since the rescattering
is very sensitive to this variable since it controls the initial particle density, e.g. for $\tau=1$ fm/c all of these effects would already
be greatly reduced \cite{Humanic:2006ib}.
The price to be paid
for this simplicity is to assume that either hadrons or ``hadron-like'' objects can exist in the
earliest stage of the heavy-ion collision just after the two nuclei pass through each other, i.e. that the hadronization time in the frame of the particle is short and 
insensitive to the environment in which it finds itself.
Clearly this simple picture is an oversimplification as is seen in
the model's shortcomings from the comparisons given above with RHIC experiments.
However, some of these shortcomings could possibly be improved by ``perturbatively'' adding a few extra pieces of physics to this model, but thereby making it less simple.

\begin{acknowledgments}
The author wishes to acknowledge financial support from the U.S.
National Science Foundation under grant PHY-0653432, and to acknowledge computing
support from the Ohio Supercomputing Center. The author also wishes to acknowledge
enlightening discussions related to this work with Hans B{\o}ggild, Ole Hansen and Boris Tomasik.
\end{acknowledgments}

\end{document}